\documentclass[aps,showpacs,twocolumn,amsmath,amssymb,nofootinbib]{revtex4-1}
\usepackage{graphicx,epsfig}
\pdfoutput=1
\usepackage{amsfonts}
\usepackage{mathrsfs}
\usepackage{multirow}
\usepackage{color}
\usepackage{graphicx}  

\def\Journal#1#2#3#4{{#1} {\bf #2}, #3 (#4)}

\def\EPJC{{Eur. Phys. J.} C}
\def\HPA{Helv. Phys. Acta}

\def\PLB{{Phys. Lett.}  B}
\def\PLA{{Phys. Lett.}  A}
\def\PRL{Phys. Rev. Lett.}
\def\PRD{{Phys. Rev.} D}
\def\PR{{Phys. Rev.}}

\def\CQG{{Class. Quant. Grav.}}
\def\CMP{{Commun. Math. Phys.}}

\def\JPA{{J. Phys.}  A}

\def\GRG{{Gen Relativ Gravit}}

\def\ibid{{\it ibid.}}

\newcommand{\be}{\begin{equation}}
\newcommand{\ee}{\end{equation}}
\newcommand{\bea}{\begin{eqnarray}}
\newcommand{\eea}{\end{eqnarray}}

\begin{document}

\title{Regular scalar clouds around a Kerr-Newman black hole: subextremal and extremal scenarios}

\date{\today}


\author{Gustavo Garc\'ia}
\email{gustavo.garcia@correo.nucleares.unam.mx} 
\author{Marcelo Salgado}
\email{marcelo@nucleares.unam.mx} \affiliation{Instituto de Ciencias
Nucleares, Universidad Nacional Aut\'onoma de M\'exico,\\
 A.P. 70-543, CDMX 04510, M\'exico}

\begin{abstract}
In this work we analyze the existence 
of electrically charged scalar clouds which are bound states of a complex-valued massive scalar field in the background of subextremal and extremal  Kerr-Newman black holes (BH). 
In particular, we reanalyze neutral (uncharged) clouds in extremal Kerr BH's.
For the extremal scenarios we have implemented a novel technique which allows us to obtain regular clouds at the BH horizon ${\cal H}^+$ which turn out to be connected ``continuously'' with the cloud solutions in the subextremal case
even if some derivatives of the scalar field are unbounded at the horizon. In particular, for subextremal BH's we have established regularity conditions at ${\cal H}^+$, by demanding that the field and its radial derivatives are bounded there, but in the extremal scenarios we relax this last condition while demanding that some scalar invariants are well behaved at ${\cal H}^+$. Furthermore we have implemented an integral technique to understand and justify in a simple and heuristic way the existence of such cloud configurations in those BH backgrounds. 
\end{abstract}

 \maketitle

\section{Introduction}
\label{Introduction}

Black holes (BH) are one of the most enigmatic and interesting predictions 
of Einstein's theory of general relativity (GR). Nowadays these objects are considered to be beyond the realm of theoretical speculations as currently we count with strong observational evidence about its presence in the universe, namely, from  the detection of gravitational waves by the LIGO-VIRGO-KAGRA collaboration~\cite{LIGOVirgo} which are emitted, among other sources, by the inspirilling and collision of two BH's, and also from the observation of the shadows produced by massive black holes at the center of the galaxies M87 \cite{EHTM87} and the Milky Way \cite{EHTSagA}, respectively. 

From a theoretical perspective, BH's seem to be very simple objects, since according to the uniqueness theorems  \cite{Uniqueness,HeuslerBook} together with the no-hair conjecture \cite{Carter1968,Ruffini1971}, they can be  described by only three parameters:  mass ($M$), angular momentum ($J$) and electric charge ($Q$) \cite{Carter1968,Ruffini1971}. More specifically, those theorems establish that all regular, stationary, axisymmetric and asymptotically flat BH solutions (AFBH) of electrovacuum Einstein's field equations belong to the Kerr-Newman family of solutions which are characterized by those three parameters.

Furthermore, these mathematical results and conjectures are supported by several {\it no-hair} theorems showing that several non-trivial fields cannot be present outside a BH. In particular, 
when one considers the Einstein-Klein-Gordon (EKG) system,  these theorems show that  
if some energy conditions are satisfied \cite{Bekenstein1972,Bekenstein1995,Sudarsky1995,Pena1997,Sudarsky1998}, the only possible scalar field  $\Psi$ (real or complex) solution present outside a static and spherically symmetric  AFBH is the trivial one $\Psi \equiv 0$, and so the only AFBH solution is the Schwarzschild solution.

Remarkably, this state of affairs can change dramatically if some spacetime symmetries are dropped and/or the energy conditions are abandoned in those theorems. For instance, when one introduces rotation (while keeping the energy conditions) it is possible to find non-trivial solutions for a 
complex-valued scalar field outside a BH. For instance, Hod was perhaps the first to show the existence of \textit{bound state} (exact)  solutions for massive but otherwise free scalar fields of that kind in the background of extremal and near extremal Kerr BH's \cite{Hod2012,Hod2013}. 
 Since the background was fixed and the scalar field was a  \textit{test field} that does not backreact in the Kerr spacetime, such solutions were termed {\it scalar clouds} and were not considered as genuine hairy solutions, i.e., solutions of the full EKG system.
 Later, Hod himself extended his results by finding exact nontrivial electrically charged scalar-cloud solutions outside (extremal and near-extremal) Kerr-Newman BH's (KNBH) \cite{Hod2014,Hod2015}. Presumably motivated by Hod's discoveries, 
 Herdeiro \& Radu \cite{Herdeiro2014,Herdeiro2015} 
 (hereafter referred to as HR)
 generalized those results in several instances. First, they found numerically scalar clouds in the background of {\it subextremal} Kerr BH's, and then, more importantly, they computed numerically genuine rotating, stationary, axisymmetric and AFBH hairy solutions by taking into account the backreaction of the field in the spacetime and thus, solving self-consistently the full EKG system under suitable regularity conditions at the horizon ${\cal H}^+$. Moreover, they showed that those hairy solutions are continuously connected with rotating {\it boson star} solutions in the limit when the BH horizon shrink to `zero'.
 
  Benone et al. \cite{Benone2014} extended Hod and HR cloud analysis by studying electrically charged scalar-field clouds in the background of subextremal KNBH. More recently, 
  we also computed scalar clouds and  rotating hairy AFBH solutions by solving numerically the full EKG system using spectral methods \cite{Grandclement2022,Garcia2023}, and confirmed the results reported by HR.
  
Delgado et al. \cite{Delgado2016} obtained hairy rotating charged AFBH solutions by solving numerically the Einstein-Klein-Gordon-Maxwell system, which extends the charged scalar-cloud analysis in a Kerr-Newman spacetime performed earlier by Benone et al \cite{Benone2014}. All those BH solutions with a non-trivial scalar hair represent some counterexamples to the no-hair conjecture.

Finally, let us mention that previously we analyzed the existence 
of clouds \cite{Garcia2019} and lack thereof \cite{Garcia2021} in the background of Kerr and 
Reissner-Nordstrom BH's, respectively, 
by devising an integral method that is usually employed in proving the no-hair theorems within, but not exclusively,  spherically symmetric scenarios. This technique provides a heuristic understanding as to why the no-scalar-hair theorems in spherically symmetric and static spacetimes cannot be extended to the axisymmetric and rotating scenarios that leads precisely to the existence of cloud and hairy solutions of the sort alluded above.

In this paper we extend the results of 
Refs.\cite{Garcia2019,Garcia2021} in several respects. First, we analyze the existence of charged scalar clouds in the background of subextremal KNBH, 
and then we study those clouds in the extremal case by introducing a novel technique that allows us to impose rigorous regularity conditions at the BH horizon ${\cal H}^+$ even when working with Boyer-Lindquist (BL) type of coordinates that are inherently singular at the horizon. In particular, we reanalyze the uncharged (neutral) clouds in the extremal Kerr background.
This novel treatment contrasts drastically with a similar analysis
performed previously by us in the extremal Kerr spacetime \cite{Garcia2020} where {\it superregularity} conditions on the scalar-field were considered by imposing boundedness in the radial derivatives of the scalar field at ${\cal H}^+$. Those conditions, while reasonable, were not necessary and led to inconsistencies that could only be remedy if the ``quantum numbers" $(n,l,m)$ that label the cloud solutions satisfied some Diophantine equation of Pell type and for a large number of nodes in the radial part of the field (i.e. $1\ll n$) \cite{Garcia2020}. In this paper we show instead that the radial derivatives of the scalar field with respect to the BL coordinate $r$, notably the first radial derivative, can be unbounded at ${\cal H}^+$ and  still leads to genuinely regular scalar-clouds  without the need of imposing any further conditions on the ``quantum numbers" $(n,l,m)$  provided that the divergence in the radial derivative is such that some invariant scalars formed from first derivatives (e.g. the {\it kinetic} term for the field in the Lagrangian) remain bounded, notably at ${\cal H}^+$.

As stressed above, this technique is applied to (electrically) charged cloud solutions in the background of extremal KN spacetime as well. We check that our solutions using that technique are in agreement with Hod's exact solutions in those two extremal backgrounds \cite{Hod2012,Hod2014}. Finally, we show that those regular cloud solutions around exact extremal Kerr and KN BH's, unlike the {\it  superregular} clouds in such extremal scenarios (cf. Ref.\cite{Garcia2020}), can be connected continuously with the cloud solutions around the corresponding subextremal BH's in the limit of extremality.
This feature hints towards the robustness of our method.

\section{Charged scalar clouds}
\label{chargedclouds}
In our study we consider a (test) massive, complex and charged scalar field $\Psi$ around a KNBH, which in Boyer-Lindquist coordinates is described by the following spacetime metric,

\begin{eqnarray}\label{metricKN}
\nonumber ds^2 = &-& \left(\frac{\Delta - a^2\sin^2\theta}{\rho^2}\right)dt^2 + \frac{\rho^2}{\Delta}dr^2 + \rho^2d\theta^2 \\  
\nonumber &-&\frac{2a\sin^2\theta\left(r^2 + a^2 - \Delta\right)}{\rho^2}dtd\varphi \\
&+& \left(\frac{\left(r^2 + a^2\right)^2 - \Delta a^2\sin^2\theta}{\rho^2}\right)\sin^2\theta d\varphi^2,
\end{eqnarray}
where
\begin{equation}
\rho^2 = r^2 + a^2\cos^2\theta\;\;, \text{and}\;\; \Delta = r^2 - 2Mr + a^2 + Q^2\;,
\label{DeltaKN}
\end{equation}
where $M$ is the mass, $a$ the angular momentum per mass unit and $Q$ the electric charge associated with the KNBH. 

In this spacetime we can identify the presence of two horizons located at
\begin{equation}\label{horizonKN}
r_{\pm} = M \pm \sqrt{M^2 - a^2 - Q^2}\;,
\end{equation}
one at $r_+ \equiv r_H$ that corresponds to the BH event horizon and another one at $r_-$ which is  an inner  Cauchy horizon, such that $\Delta (r_{\pm})=0$
. The existence of a KNBH requires $a^2 + Q^2 \leq M^2$, where the equality is associated with an {\it extremal} KNBH, which we consider in Sec.~\ref{Extremalscenario}.
In particular, the Kerr spacetime is recovered when $Q\equiv 0$.
Due to the presence of the two horizons, it is convenient to write
\begin{equation}
\Delta = (r - r_H)(r - r_-)\;,
\end{equation}
where the values of $r_H$ and $r_-$ keep the following relationship
\begin{equation}\label{CauchyHorizon}
r_- = \frac{a^2 + Q^2}{r_H}\;.
\end{equation}
The angular velocity of the KNBH is given as follows:
\begin{equation}\label{OmegaHKN}
\Omega_H = \frac{a}{r_H^2 + a^2}\;.
\end{equation}

From (\ref{DeltaKN}) and the fact that 
$\Delta (r_H)=0$ one finds a relation between the mass $M$ and the quantities $r_H$, $a$ and $Q$:
\begin{equation}\label{MassKN}
M = \frac{r_H^2 + a^2 + Q^2}{2r_H}\;.
\end{equation}

Moreover, from Eqs.(\ref{CauchyHorizon})--(\ref{MassKN}) we see that the quantities $r_-$, $\Omega_H$, and $M$ can take a parametric  form $r_- = r_-(r_H, a, Q)$, $\Omega_H = \Omega_H(r_H, a, Q)$, and $M = M(r_H, a, Q)$. These equations will allow us to compute $r_-$, $\Omega_H$, and $M$ when finding the values for $a$ that solve the eigenvalue problem for $\Psi$, once the values of $r_H$ and $Q$ are specified (cf. Sec.~\ref{Teukolsky}).

The massive and charged scalar field $\Psi$ that we analyze for the existence of scalar-cloud solutions has the following energy-momentum tensor (EMT):
\begin{eqnarray}
\label{TabKN}
\nonumber T_{ab} &=& \frac{1}{2}\Big[\left(D_{a}\Psi\right)^*\left(D_{b}\Psi\right) + \left(D_{b}\Psi\right)^*\left(D_{a}\Psi\right)\Big] \\
&-& g_{ab}\Big[\frac{1}{2}g^{cd}\left(D_{c}\Psi\right)^*\left(D_{d}\Psi\right) + U(\Psi^*\Psi)\Big]\;, 
\end{eqnarray}
where the operator 
\begin{equation}\label{operatorKN}
D_a \equiv \nabla_a - iqA_a\;,
\end{equation}
represents the covariant derivative associated with the \textit{gauge} field $A_a$, which in the KNBH background is given by
\begin{equation}\label{4potentialKN}
A_a = -\frac{Qr}{\rho^2}\left[(dt)_a - a\sin^2\theta(d\varphi)_a\right]\;,
\end{equation}
and the constant $q$ (i.e., the \textit{electrical charge}) is the gauge coupling for the scalar field $\Psi$. The operator $\nabla_a$ corresponds to the covariant derivative compatible with the metric, in this case the KN metric.
For our study we focus only on the following potential.
\begin{equation}\label{PotentialBH}
U(\Psi^*\Psi) = \frac{1}{2}\mu^2 \Psi^*\Psi\;,
\end{equation}
which is associated with a massive but \textit{free field} with mass $\mu$.

The dynamics of the charged and massive scalar field is given by the Klein-Gordon (KG) equation coupled to the electromagnetic potential:
\begin{equation}\label{KGequationKN}
\left(\nabla^{a} - iqA^{a}\right)\left(\nabla_a - iqA_a\right)\Psi = \mu^2\Psi\;.
\end{equation}

In order to find \textit{bound states} for $\Psi$ in the domain of outer communication (DOC) of the KNBH,
including the horizon, we consider the following ansatz in terms of the BL coordinates with temporal and angular dependence in the form, 
\begin{equation}
\label{chargedfieldKN}
\Psi(t, r, \theta, \varphi) = \phi(r, \theta)e^{im\varphi}e^{-i\omega t}\;,
\end{equation} 
where $\omega$ is the frequency of the scalar field and $m$ is an integer number. This is the most general form that we can choose in such a way that the energy-momentum tensor (\ref{TabKN}) respects the symmetries of the KN spacetime.

To ensure the existence of boson clouds, we impose the \textit{zero flux condition} at the BH horizon \cite{Herdeiro2014,Herdeiro2015}:
\begin{equation}
  \label{ZFC} 
\chi^aD_a\Psi\big|_{\mathcal{H}^+} = 0;,
\end{equation}
where 
\begin{equation}\label{HellicalK}
\chi^a \equiv \xi^a + \Omega_H\eta^a\;,
\end{equation}
is the \textit{helical} Killing vector field given in terms of the timelike Killing field $\xi^a = (\partial/\partial t)^a$ and the axial Killing field $\eta^a = (\partial/\partial\varphi)^a$, which are associated with the time and axial symmetries of the
background spacetime. At the horizon $\chi^a$ becomes null and thus, it is tangent to the null geodesic generators of the horizon. From (\ref{ZFC})  together with Eqs.(\ref{HellicalK}), (\ref{4potentialKN}) and (\ref{chargedfieldKN}), we obtain the following condition:
\begin{equation}
\label{Cond1}
\left(\omega - m\Omega_H - q\frac{Qr_H}{r_H^2 + a^2}\right)\Psi_H = 0
\end{equation}
Assuming that in general $\Psi_H \neq 0$, we conclude,
\begin{equation}
  \label{synchronicity}
\omega = m\Omega_H + q\Phi_H\;,
\end{equation}
where 
\begin{equation}
\Phi_H := -A_a \chi^a |_{\mathcal{H}^+}= \frac{Qr_H}{r_H^2 + a^2}=\frac{4\pi Qr_H}{{\cal A}_H} \;,
\end{equation}
is the electric potential at the horizon as defined in terms of the helical Killing field, and 
${\cal A}_H= 4\pi (r_H^2 + a^2)$ is the area of the BH event horizon (cf. Ref.~\cite{Hawking1976}).

The bound states thus correspond to a field with a frequency given by (\ref{synchronicity}), which is called \textit{synchronicity condition} \cite{Hod2014,Benone2014}, due to its relationship with the BH's angular velocity $\Omega_H$, in particular when $Q\equiv 0$.

In fact, the condition 
(\ref{synchronicity}) results also when imposing regularity of the field $\Psi$ at the horizon, as we show below.

\section{Teukolsky radial equation and regularity conditions}
\label{Teukolsky}

When we substitute the scalar field \textit{ansatz} (\ref{chargedfieldKN}) into the KG Eq.(\ref{KGequationKN}) we find that the function $\phi(r, \theta )$ is separable, that is, it can be written as the product of a function dependent on the variable $r$ and a function dependent on the variable $\theta$ which allows us to expand the $\Psi$ field in modes of the form
\begin{equation}\label{ScalarFieldKN}
\Psi_{nlm} = R_{nlm}(r)S_{lm}(\theta)e^{im\varphi}e^{-i\omega t}\;,
\end{equation}
where the angular functions $S_{lm}\left(\theta\right)$ 
(the \textit{spheroidal harmonics}) obey the following equation 
\begin{eqnarray}
\nonumber  
&&\!\!\!\!\!\!\!\!\frac{1}{\sin\theta}\frac{d}{d\theta}\left(\sin\theta\frac{dS_{lm}}{d\theta}\right) \\
  && \!\!\!\!\!\!\!\! + \left(K_{ lm} + a^2(\mu^2 - \omega^2)\sin^2\theta - \frac{m^2}{\sin^2\theta}\right)S_{lm} = 0 \,,\label{SangularKN}
\end{eqnarray}
and $K_{lm}$ are the separation constants ($|m| \leq l$) given by 
\begin{equation}
  \label{KlmKN}
K_{lm} + a^2(\mu^2 - \omega^2) = l(l + 1) + \sum_{k=1}^{\infty}c_{k}a^{2k}(\mu^2 - \omega^2)^k\;,
\end{equation}
which connects the angular and radial parts of the KG equation and ensures that the angular functions $S_{lm}(\theta)$ are regular on the axis of symmetry. 
As it occurs with the wave function of the hydrogen atom, 
the number $l$ is a non-negative integer which is associated with the angular momentum of the atom, while the integer $n$ ($n\geq 0$) labels the number of {\it nodes} in the radial function $R_{nlm}(r)$. The expansion coefficients $c_k$ can be found in Ref.~\cite{Abramowitz}.

The functions $R_{nlm}(r)$ obey a type of radial Teukolsky equation \cite{Teukolsky1972},
\begin{eqnarray}
\nonumber && \Delta\frac{d}{dr}\left(\Delta\frac{dR_{nlm}}{dr}\right) + \left[\mathcal{H}^2 + \left(2ma\omega - K_{lm}\right.\right. \\
&-& \left.\left.\mu^2\left(r^2 + a^2\right)\right)\Delta\right]R_{nlm} = 0\;,\label{Teukolsky-KN}
\end{eqnarray}  
where
\begin{equation}\label{HKN}
\mathcal{H} \equiv \left(r^2 + a^2\right)\omega - am - qQr\;.
\end{equation}
Given the form of the frequency $\omega$ (\ref{synchronicity}), we observe that the function $\mathcal{H}$ vanishes at the horizon, which is precisely the regularity condition at the horizon when demanding boundedness for the field and its radial derivatives in the subextremal BH scenarios, as we discuss next.

In order to find cloud configurations, it is essential to establish regularity conditions for the $\Psi$ field at the event horizon, namely, by demanding that 
the field and some of its derivatives are bounded on the horizon. We thus  assume that $R_{nlm}(r)$ is $C^3$ on the horizon
. To solve the differential equation associated with the radial part (\ref{Teukolsky-KN}) it is necessary to know the values 
of the field and its  radial derivatives at $r_H$: $R''_{nlm}(r_H)$, $R'_{nlm}(r_H) $ and $R_{nlm}(r_H)$. So, assuming that $R''_{nlm}(r_H)$ is bounded we find the regularity condition for $R'_{nlm}(r_H)$ from Eq.(\ref{Teukolsky-KN}): 
\begin{eqnarray}
\nonumber R'_{nlm}(r_{H}) &=& -\frac{1}{2(r_H - M)}\left[2ma\omega - K_{lm} - \mu^2\left(r_H^2 + a^2\right)\right]\\
&& \times R_{nlm}(r_H)\;.\label{RCKN1}
\end{eqnarray}
The value of $R_{nlm}(r_H)$ is a free parameter, which we take $R_{nlm}(r_H) = 1$, for simplicity  and also to compare our numerical results with those presented in \cite{Benone2014} where the same value is used.
To find the value for $R''_{nlm}(r_H)$, we differentiate Eq. (\ref{Teukolsky-KN}) once more, and demand that $R'''_{nlm}(r_H)$ is bounded, which leads to\footnote{We can observe that, for $Q = 0$ and $a \neq 0$ these regularity conditions reduce to those obtained in the case of the Kerr metric (see Eqs.~(77)-(78) in \cite{Garcia2019}). Alternatively, when $a = 0$ and $Q \neq 0$, they reduce to the regularity conditions for the field in the background of a Reissner-Nordstrom (RN) black hole (see Eqs.~(18)-(19) in \cite{Garcia2021}).}:
\begin{widetext}
\begin{eqnarray}
\nonumber R''_{nlm}(r_{H}) &=& - \frac{1}{2(r_H - M)}\left[\frac{(M - r_-)\left(2mar_H + qQ \left(r_H^2 - a^2\right)\right)^2}{(r_H - r_-)^2\left(r_H^2 + a^2\right)^2} - \mu^2r_H\right]R_{nlm}(r_H)\\
&-&\frac{1}{4(r_H - M)}\left[2(1 + ma\omega) - K_{lm} - \mu^2\left(r_H^2 + a^2\right)\right]R'_{nlm}(r_{H})\;.\label{RCKN2}
\end{eqnarray}
\end{widetext}
Both regularity conditions (\ref{RCKN1}) and (\ref{RCKN2})
are valid only in the subextremal case $r_H\neq M$. 
Thus, making use of these two regularity conditions we can find cloud solutions close to the extremal background {\it only} in the limit $r_H \rightarrow M$, but we cannot treat  exactly the extremal scenario with such conditions. 
Clouds in the exact extremal background with $r_H=M=r_-$ will be analyzed separately in Sec.\ref{Extremalscenario} below.

In order to find charged cloud solutions around a subextremal KNBH Eq.(\ref{Teukolsky-KN}) was solved numerically with their respective regularity conditions (\ref{RCKN1}) and (\ref{RCKN2}), using a 4th order Runge-Kutta algorithm and integrating in the DOC from $r = r_H$ outwards. As a part of the numerical scheme, the values of $r_H$, $Q$, $\mu$ and $q$ were fixed, and selecting some values for the integers $n$, $l$ and $m$. Then the BH angular momentum per mass unit $a$ can be used as an {\it eigenvalue} such that the radial function $R_{nlm}(r)$ vanishes asymptotically as we did in our previous study for clouds around a subextremal Kerr BH \cite{Garcia2019}.

\begin{figure}[h!]
\begin{center}
\includegraphics[width=0.53\textwidth]{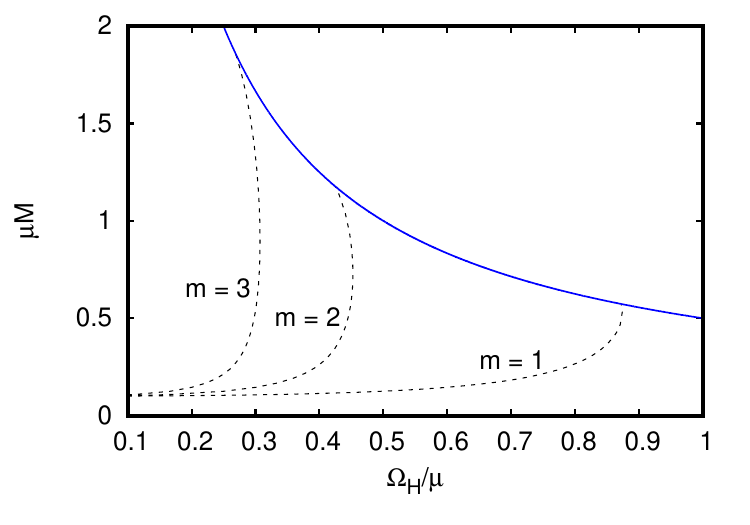}
\caption{Existence (black dotted) lines for charged scalar clouds ($n=0$ and $l=m$ with $m=1, 2, 3$) coupled to a subextremal Kerr-Newman black hole with charge $\mu Q = 0.1$. The relation between the mass and the charge of the scalar field is $q/\mu = 1$. The  solid blue line corresponds to an extremal KNBH  $a^2 + Q^2 = M^2$ (BH solutions do not exist above this line).}
\label{fig:RadialKN1}
\end{center}
\end{figure}

Figure~\ref{fig:RadialKN1} shows the \textit{existence lines} associated with the numerical solutions of Eq.(\ref{Teukolsky-KN}), with fixed charge $\mu Q = 0.1$ and different values for $r_H$. We have taken the \textit{quantum numbers} $n = 0$ and $m = l = 1, 2, 3$, respectively, as well as a scalar field $\Psi$ with $q/\mu = 1$.

In Tables~\ref{tab:spectrum1KN}-\ref{tab:spectrum3KN} of  Appendix~\ref{sec:appendixKN} are displayed the numerical eigenvalues $a_{nlm}$  (i.e., the \textit{spectrum}) associated with the eigenvalue problem of Eq.(\ref{Teukolsky-KN}). Our results depicted in Figure~\ref{fig:RadialKN1} are consistent with the those reported in \cite{Benone2014}.

\begin{figure}[h!]
\begin{center}
\includegraphics[width=0.5\textwidth]{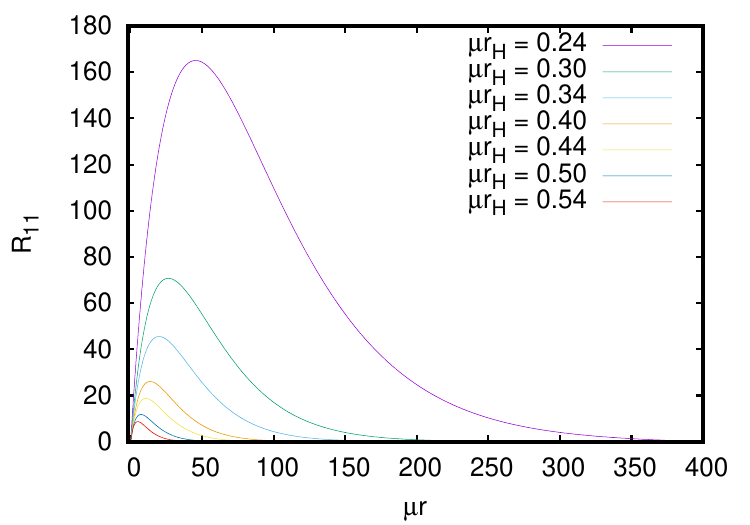}
\caption{Radial solutions $R_{11}$ ($R_{011}$) with $n = 0$ and $l = 1 = m$ associated with charged boson clouds around a subextremal Kerr-Newman black hole and the corresponding location of the event horizon $\mu r_{H}$ taking $\mu Q = 0.1$ and $q/\mu = 1.0$.}
\label{fig:RadialKN2}
\end{center}
\end{figure}

Figure~\ref{fig:RadialKN2} plots a family of radial solutions $R_{nlm}(r)$ with $n = 0$ and $m = l = 1$, using different values for $r_H$. The  corresponding values for the spectrum $a_{nlm}$ are shown in Table~\ref{tab:spectrum1KN} of  Appendix~\ref{sec:appendixKN}.

\begin{figure}[h!]
\begin{center}
\includegraphics[width=0.5\textwidth]{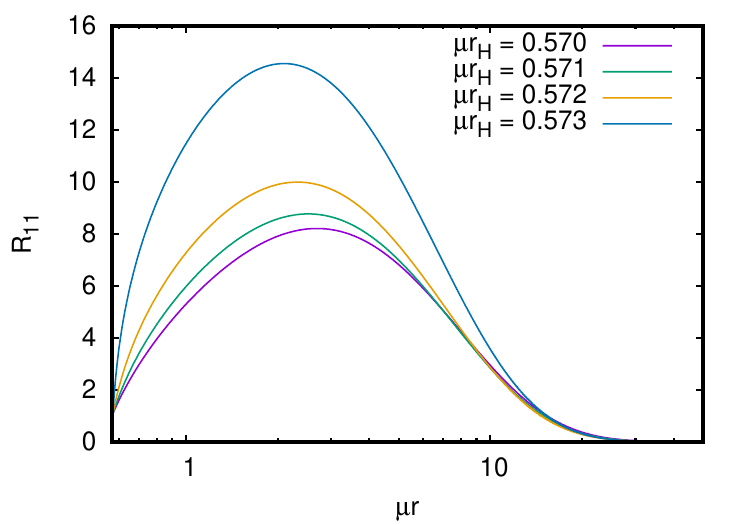}
\caption{Radial solutions $R_{11}$ ($R_{011}$) with $n = 0$ and $l = 1 = m$ associated with charged boson clouds around a Kerr-Newman black hole close to extremality $r_H \approx M$. Here $\mu Q = 0.1$ and $q/\mu = 1.0$ are fixed.}
\label{fig:RadialKN3}
\end{center}
\end{figure}

Cloud solutions around KNBH close to extremality $a^2 + Q^2 \approx M^2$ (i.e. $r_H \approx M$) are depicted in  Fig.~\ref{fig:RadialKN3} for the fundamental mode ($n = 0$) and $m = l = 1$. The numerical values for the spectrum $a_{nlm}$
can be found in the last four rows and second column of Table~\ref{tab:spectrum1KN} (see Appendix~\ref{sec:appendixKN}).
For clouds similar to these ones,
Tables~\ref{tab:spectrum2KN} and \ref{tab:spectrum3KN} display
more values $a_{nlm}$  when $m = l = 2$ and $m = l = 3$, respectively. Note from Fig.~\ref{fig:RadialKN3} that as
the solutions approach the extremal scenario the slope $R'$ at the horizon increases (cf. 
Eq.(\ref{RCKN1}) ). At this respect, cloud solutions for the {\it exact} extremal scenario around
Kerr and Kerr-Newman BH's are analyzed below in Sec.\ref{Extremalscenario} since, as stressed before, the radial part $R(r)$
for the boson field $\Psi$ requires different kind of regularity conditions at the horizon because
their derivatives (\ref{RCKN1}) and (\ref{RCKN2}) blows-up when $r_H=M$.

\begin{figure}[h!]
  \centering
   \label{fig:KNn11}
    \includegraphics[width=0.5\textwidth]{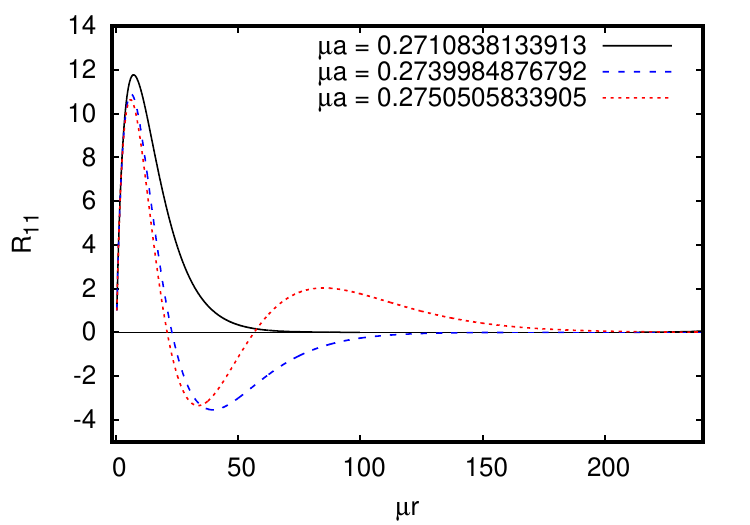}
    \label{fig:KNn22}
     \includegraphics[width=0.5\textwidth]{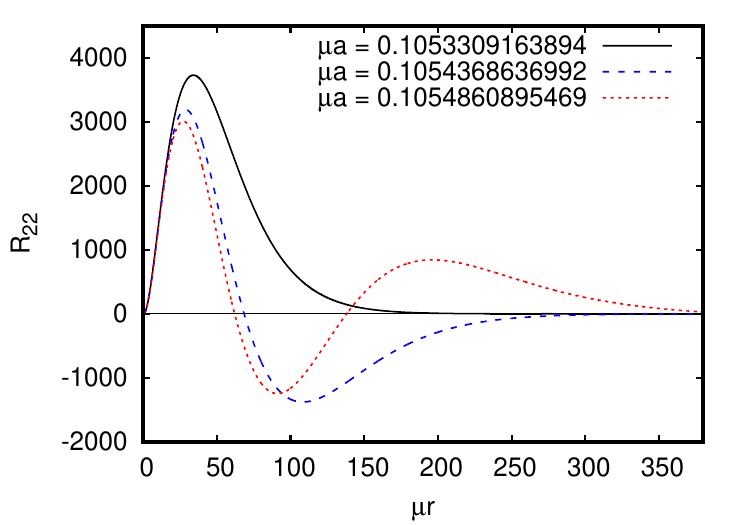}
     \caption{Radial solutions $R_{nlm}$ with principal number $n = 0, 1, 2$ (number of nodes) around
       a KNBH with the horizon located at $\mu r_H = 0.5$, and taking integers $m = l = 1$ (top panel) and $m = l = 2$ (bottom panel), respectively. Here $\mu Q = 0.1$ and $q/\mu = 1.0$.}\label{fig:Radialpart12}
\end{figure}

Figure~\ref{fig:Radialpart12} shows some examples of radial solutions  $R_{nlm}$ with different nodes
($n = 0, 1, 2$) associated with a fixed value of the horizon radius $\mu r_H = 0.5$. In particular, the figures
depicts $R_{n11}$ and $R_{n22}$, i.e.,  taking $m = l = 1$ and $m = l = 2$, respectively.

\begin{figure}[h!]
  \centering
   \label{fig:S22KN}
    \includegraphics[width=0.5\textwidth]{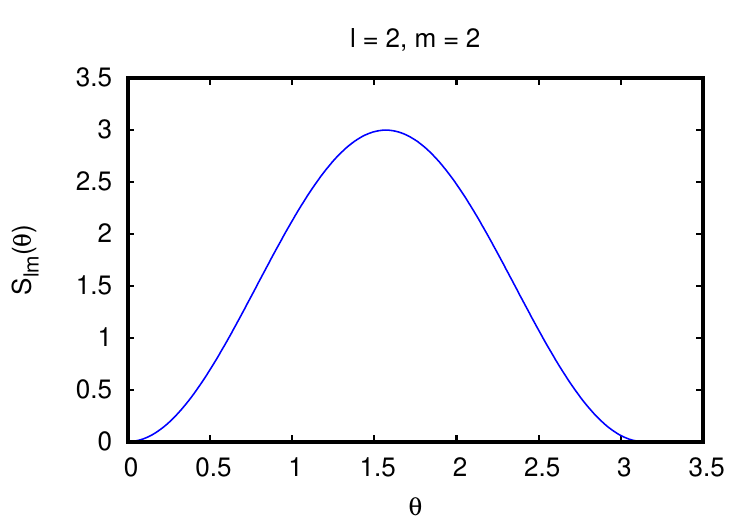}
    \label{fig:SS22KN}
     \includegraphics[width=0.5\textwidth]{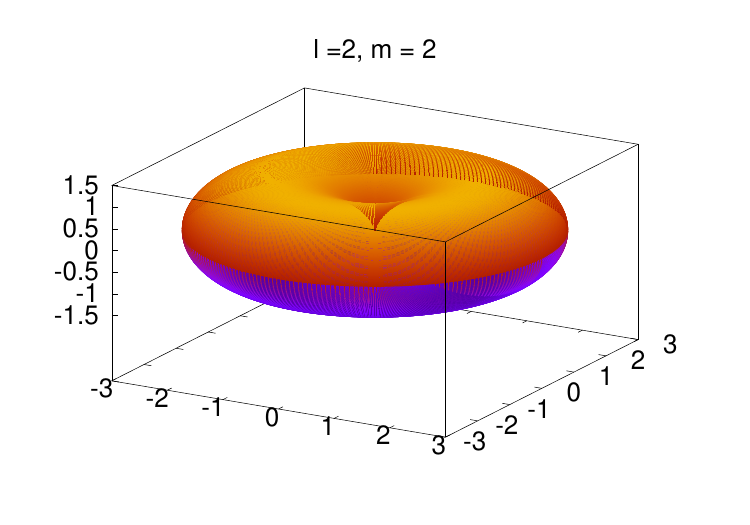}
\caption{Angular part $S_{lm}(\theta)$ of the field $\Psi$ (top panel) and 3D spheroidal harmonic $|S_{22}(\theta,\varphi)|$ (bottom panel) for $m = l = $2.}\label{fig:Sm22KN}
\end{figure}

Figure~\ref{fig:Sm22KN} shows the angular function $S_{lm}(\theta)$ and the 3D spheroidal harmonic $|S_{lm}(\theta,\varphi)|$ for $m = l = 2$, associated with the charged scalar field $\Psi$ coupled to a Kerr-Newman black hole with $\mu r_H = 0.5$ and $\mu Q = 0.1$.

\section{No-hair-theorem obstructions}
\label{NoHairtheorem}

In this Section we justify in a more heuristic way the existence of charged scalar clouds in the presence of a KNBH. The treatment developed here is similar to the one presented in \cite{Garcia2019,Garcia2021}, which analyze, respectively,
the existence of neutral scalar
clouds around a Kerr BH, and their absence thereof around a RNBH. To do so, we
use a technique initially implemented by Bekenstein \cite{Bekenstein1972}, and then employed, {\it mutatis mutandis},
in several studies that prove no-hair theorems in several kind of scalar-field theories \cite{Heusler1996,Ayon2002}.

First, let us consider the Klein-Gordon equation (\ref{KGequationKN}) in the form
\begin{equation}\label{KGequationKN2}
D^aD_a\Psi = \mu^2\Psi\;.
\end{equation}
Multiplying both sides of this equation by $\Psi^*$ and integrating over a suitable volume $\mathcal{V}$ contained within the DOC
of the KNBH (bounded by two spatial hypersurfaces $t=const$, a section of the horizon and a timelike surface
$r=const$ when $r\rightarrow \infty$) we find,
\begin{equation}\label{IntegralKN1}
\int_\mathcal{V}\Psi^*D^aD_a\Psi\sqrt{-g}d^{4}x = \int_\mathcal{V}\mu^2\Psi^*\Psi\sqrt{-g}d^{4}x\;.
\end{equation}

Integrating by parts the l.h.s of the previous equation and using the Gauss theorem we obtain,
\begin{equation}
\int_{\partial\mathcal{V}}\Psi^*s^aD_a\Psi dS = \int_\mathcal{V}\Big[(D^a \Psi)^*(D_a \Psi) + \mu^2\Psi^*\Psi\Big]\sqrt{-g}d^{4}x
\end{equation}

In this case we have that the surface integral that correspond to the boundary $\partial \mathcal{V}$ is composed by four regions: the two spacelike hypersurfaces $\Sigma_1$ and $\Sigma_2$, a portion of the black hole horizon and, finally, 
the asymptotic region corresponding to spatial infinity $i^0$. The contributions associated with $\Sigma_1$ and $\Sigma_2$ cancel each other because the integrals are identical except for the fact that their normal vectors have opposite signs (i.e. $s^a_{\Sigma_1} = - s^a_{\Sigma_2}$).
The surface integral associated with spatial infinity vanishes when assuming that the scalar field $\Psi$ decays exponentially,
and thus it vanishes asymptotically. Finally, at the horizon (a null hypersurface) the normal $s^a$ is equal to the Killing vector $\chi^a$. Furthermore, assuming that $\Psi^*$ is bounded at the horizon and using the zero-flux condition (\ref{ZFC})
the integral associated with the portion of horizon vanishes as well. We conclude that 
\begin{equation}
\int_{\partial\mathcal{V}}\Psi^*s^aD_a\Psi dS = 0\;,
\end{equation}
and therefore
\begin{equation}
\label{IntegralKN}
\int_\mathcal{V}\Big[(D^a \Psi)^*(D_a \Psi) + \mu^2\Psi^*\Psi\Big]\sqrt{-g}d^{4}x =0 \,.
\end{equation}

Given the harmonic dependence of the field $\Psi$ (\ref{chargedfieldKN}) one obtains the following expression for the \textit{kinetic term}
\begin{eqnarray}
\label{KineticKN}
\nonumber K &\equiv& \left(D^a\Psi\right)^*\left(D_a\Psi\right) \\
\nonumber &=& g^{tt}\left(D_t\Psi\right)^*\left(D_t\Psi\right) + g^{t\varphi}\left(D_t\Psi\right)^*\left(D_{\varphi}\Psi\right) \\
\nonumber &+& g^{\varphi t}\left(D_{\varphi}\Psi\right)^*\left(D_t\Psi\right) + g^{\varphi\varphi}\left(D_{\varphi}\Psi\right)^*\left(D_{\varphi}\Psi\right) \\ 
\nonumber &+& g^{rr}\left(D_{r}\Psi\right)^*\left(D_{r}\Psi\right) + g^{\theta\theta}\left(D_{\theta}\Psi\right)^*\left(D_{\theta}\Psi\right)\\
&=& \mathcal{R}\phi^2 + g^{rr}\left(\partial_r\phi\right)^2 + g^{\theta\theta}\left(\partial_{\theta}\phi\right)^2\;,
\end{eqnarray}
where we have defined
\begin{eqnarray}
  \label{R-KN}
  \mathcal{R} &\equiv& g^{tt}\hat{\omega}^2 - 2g^{t\varphi}\hat{m}\hat{\omega} + g^{\varphi\varphi}\hat{m}^2\;, \\
  \hat{\omega} &\equiv& \omega + qA_t\;,
  \label{hatomega} \\
  \hat{m} &\equiv& m - qA_{\varphi}\;.
  \label{hatm}
\end{eqnarray}
Considering the frequency of the field $\Psi$ (\ref{synchronicity}) and the electromagnetic potential $A_a$ (\ref{4potentialKN}), it is possible to express Eqs.~(\ref{hatomega}) and (\ref{hatm}) as follows:
\begin{eqnarray}
  \label{hatomega2}
\hat{\omega} &=& m\Omega_H + q\Phi_H - \frac{qQr}{\rho^2}\;, \\
\label{hatm2}
\hat{m} &=& m - \frac{aqQr\sin^2\theta} {\rho^2}\;.
\end{eqnarray}

From Eqs.~(\ref{hatomega2}) and (\ref{hatm2}) together with (\ref{synchronicity})
it is easy to find the following relationship between
$\hat{\omega}$ and $\hat{m}$:
\begin{equation}
\hat{m} = m + (\hat{\omega} - \omega)a\sin^2\theta\;.
\end{equation}

If one substitutes the explicit form of the metric components $g^{tt}$, $g^{t\varphi}$, and $g^{\varphi\varphi}$ into Eq.(\ref{R-KN}), one obtains the following expression for $\mathcal{R}$:
\begin{eqnarray}
\nonumber \mathcal{R} = &-&\frac{1}{\Delta\rho^2}\left[\left(r^2 + a^2\right)\hat{\omega} - a\hat{m}\right]^2\\
 &+& \frac{1}{\rho^2\sin^2\theta}\left[a\hat{\omega}\sin^2\theta - \hat{m}\right]^2\;, 
\end{eqnarray} 
or equivalently
\begin{eqnarray}\label{RKNBL}
\nonumber \mathcal{R} = &-&\frac{(r - r_H)}{\rho^2(r - r_-)}\left[\omega r + \left(mr_H - aqQ\right)\Omega_H\right]^2 \\
&+& \frac{1}{\rho^2\sin^2\theta}\left[a\omega\sin^2\theta - m\right]^2\;.
\end{eqnarray} 
From this last equation, we observe that $\mathcal{R}$ contains a contribution that is never positive
(the one with the factor $r-r_H$ in the first line) and another contribution that is positive defined in the DOC (including
the horizon, in the second line).

Therefore, the \textit{kinetic term} $K$ is represented by the following expression:
\begin{eqnarray}
\nonumber K = &-&\frac{(r - r_H)\phi^2}{\rho^2(r - r_-)}\left[\omega r + \left(mr_H - aqQ\right)\Omega_H\right]^2 \\
\nonumber &+& \frac{\phi^2}{\rho^2}\left[\frac{a\omega\sin^2\theta - m}{\sin\theta}\right]^2\\ 
&+& g^{rr}\left(\partial_r\phi\right)^2 + g^{\theta\theta}\left(\partial_{\theta}\phi\right)^2\;.\label{KineticTermKN}
\end{eqnarray}

Therefore, the integrand of the integral (\ref{IntegralKN}) has the following form
\begin{eqnarray}
\nonumber I &\equiv& K + \mu^2\Psi^*\Psi \\
\nonumber &=& -\,\frac{(r - r_H)\phi^2}{\rho^2(r - r_-)}\left[\omega r + \left(mr_H - aqQ\right)\Omega_H\right]^2 \\
\nonumber && +\, \frac{\phi^2}{\rho^2}\left[\frac{a\omega\sin^2\theta - m}{\sin\theta}\right]^2\\
&& +\;\; \frac{\Delta}{\rho^2}\left(\partial_r\phi\right)^2 + \frac{1}{\rho^2}\left(\partial_{\theta}\phi\right)^2 + \mu^2\phi^2\;.\label{IntegrandKN}
\end{eqnarray}

If nontrivial regular charged boson clouds exist (i.e. $\phi(r, \theta) \neq 0$ in general), the following inequality
\begin{equation}\label{inequalityKN}
\mathcal{R} = g^{tt}\hat{\omega}^2 - 2g^{t\varphi}\hat{m}\hat{\omega} + g^{\varphi\varphi}\hat{m}^2 \leq 0\;,
\end{equation} 
must hold in a region of the KN spacetime so that the (non positive) term in the first line of (\ref{IntegrandKN}) compensates the
non-negative contributions associated with the quadratic terms $g^{rr}\left( \partial_r\phi\right)^2 + g^{\theta\theta}\left(\partial_{\theta}\phi\right)^2$ and $\mu^2\phi^2$, and in this way the volume integral (\ref{IntegralKN}) is satisfied for
non-trivial clouds. We show that this is the case in some particular instances, and then provide numerical evidence
in more general cases.

First, considering Eq.(\ref{RKNBL}) we find the following behaviors at the horizon $r_H$ and asymptotically
(when $r \rightarrow \infty$), respectively:
\begin{equation}
\label{RKNhorizon}
\mathcal{R}_{H} = \frac{\left[aqQr_H\sin^2\theta - \left(r_H^2 + a^2\cos^2\theta\right)m\right]^2}{(r_H^2 + a^2\cos^2\theta)(r_H^2 + a^2)^2\sin^2\theta}\;, 
\end{equation} 
\begin{equation}
\label{RKNinfty}
\mathcal{R}_{\infty} \sim  -\omega^2\;.
\end{equation}
In particular, at the equatorial plane $\theta = \pi/2$, Eq.(\ref{RKNhorizon}) reduces to
\begin{equation}\label{RKNhorizonpi2}
\mathcal{R}_{H} = \left[\frac{aqQ - mr_H}{r_H^2 + a^2}\right]^2\;. 
\end{equation}
Clearly, this shows that $\mathcal{R}$ must interpolate between a positive value at the horizon
(\ref{RKNhorizonpi2})  and a negative value at spatial infinity (\ref{RKNinfty}).

Furthermore, since in the kinetic term appears the combination $\phi^2\mathcal{R}$, we introduce the quantity $\Lambda$,
\begin{equation}\label{LambdaKN}
\Lambda \equiv \phi^2\mathcal{R}\;,
\end{equation}
which presents the following behaviors at the horizon and the equatorial plane and asymptotically ($r_H \ll r$), respectively
\begin{eqnarray}\label{LambdaKNH}
  \Lambda\big|_H &=& \phi^2_H\mathcal{R}_H = m^2\phi^2_H\left[\frac{aqQ - mr_H}{r_H^2 + a^2}\right]^2 \;,\\
  \Lambda_{\infty} &=& \phi^2_\infty\mathcal{R}_{\infty} \sim - \phi^2_\infty\left(m\Omega_H + q\Phi_H\right)^2\;.
\end{eqnarray}
In particular, $\Lambda_{\infty}$ vanishes as $\phi \rightarrow 0$ asymptotically.

On the other hand, when considering an extremal KNBH, $a^2 + Q^2 = M^2$, the quantity $\mathcal{R}$ given in Eq.(\ref{RKNBL})
reduces to 
\begin{eqnarray}\label{RextKN}
\nonumber \mathcal{R}^{\rm ext} = &-& \frac{1}{\rho^2}\left[\omega r + \left(mM - aqQ\right)\Omega_H\right]^2\\
&+& \frac{1}{\rho^2}\left[\frac{a\omega\sin^2\theta - m}{\sin\theta}\right]^2\;.
\end{eqnarray}
Then, the frequency of the scalar field is given by,
\begin{equation}\label{omegaKNext}
\omega = \frac{ma}{M^2 + a^2} + \frac{qQM}{M^2 + a^2} = \frac{ma + qQM}{M^2 + a^2}\;.
\end{equation}

In this case, the value of $\mathcal{R}^{\rm ext}$ at the horizon $r = r_H = M$ takes the form,
\begin{eqnarray}\label{RextKN2}
  \nonumber && \mathcal{R}^{\rm ext}_H = -\frac{1}{M^2 + a^2\cos^2\theta}\left\lbrace\left[\frac{qQ^3 + 2maM}{M^2 + a^2}\right]^2\right. \\
&+& \left.\left[\frac{ma^2\cos^2\theta + mM^2 - aqQM\sin^2\theta}{(M^2 + a^2)\sin\theta}\right]^2\right\rbrace \;,
\end{eqnarray} 
which, at the equatorial plane ($\theta = \pi/2$) reduces to
\begin{equation}
\mathcal{R}^{\rm ext}_H = -\frac{1}{M^2}\left[\frac{qQ^3 + 2maM}{M^2 + a^2}\right]^2 + \left[\frac{aqQ - mM}{M^2 + a^2}\right]^2\;.
\end{equation}
We note that when $Q = 0$, the above equation is simply
\begin{equation}
  \label{RextH}
\mathcal{R}^{\rm ext}_H = -\frac{3m^2}{4M^2}\;,
\end{equation}
this value is similar the one that appears in the equation (61) of \cite{Garcia2019}, and which corresponds to the extremal Kerr scenario\footnote{In \cite{Garcia2019} we use a slightly different notation and the equivalent
  expression for (\ref{RextH}) does not include the factor $m^2$.}.

The asymptotic behavior for $\mathcal{R}^{\rm ext}$ 
when $M \ll r$, coincides with the asymptotic behavior for $\mathcal{R}$ given by Eq.~(\ref{RKNinfty}).

In view of this, even if $\mathcal{R}^{\rm ext}_H$ were positive for some values of the parameters,
$\mathcal{R}^{\rm ext}$ must interpolate from a positive to a negative value, proving that this quantity is also 
negative in a region of the DOC. 

\begin{figure}[h!]
\begin{center}
\includegraphics[width=0.45\textwidth]{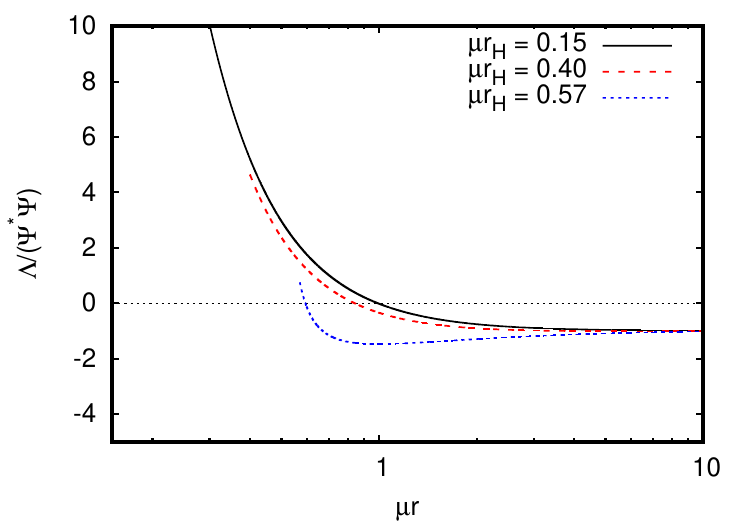}
\includegraphics[width=0.45\textwidth]{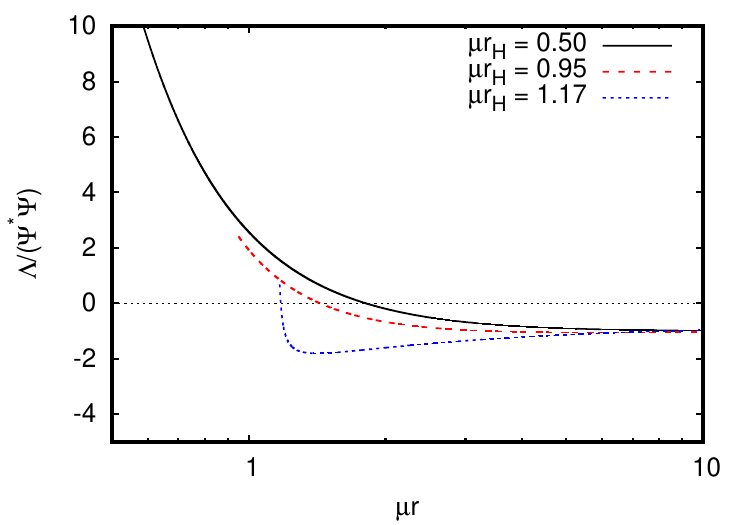}
\includegraphics[width=0.45\textwidth]{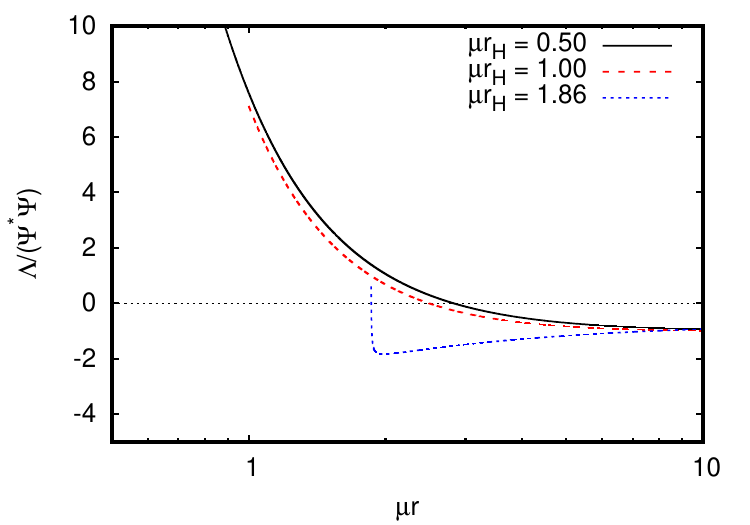}
\caption{Rotational contribution [cf. Eq.~(\ref{LambdaKN})]
  $\Lambda/\Psi^*\Psi= {\cal R}$ to the kinetic term $K= (\nabla_c \Psi^*)(\nabla^c\Psi)$
  given by Eq.~(\ref{KineticTermKN}) computed from the {\it regular} cloud solutions
  in the background of a Kerr-Newman BH and evaluated at the equatorial plane ($\theta=\pi/2$) for different locations of the horizon $r_H$, and with ``quantum numbers'' $n = 0$ and $m = l = 1 $ (top panel), $m = l = 2$ (middle panel), and $m = l = 3$ (bottom panel).}\label{fig:kineticTermKN}
\end{center}
\end{figure}

We conclude that the contribution associated with Eq.(\ref{LambdaKNH}) to the \textit{kinetic term} (\ref{KineticTermKN})
is positive at the horizon (in the subextremal case) and then it becomes negative. In the extremal
case something similar happens, except that at the horizon $\mathcal{R}^{\rm ext}$ can be even less positive
(cf. blue dashed-line of top panel of Fig.\ref{fig:kineticTermKN} which corresponds to a near extremal solution).

Since the field vanishes asymptotically, it follows that the $\Lambda$ term (in fact the entire kinetic term) will
also vanish asymptotically. As in the Kerr scenario \cite{Garcia2019}, we observe that the existence of nontrivial localized solutions for $\phi(r, \theta)$ in a Kerr-Newman background implies that the inequality (\ref{inequalityKN}) must hold in some region of the DOC.

Figure~\ref{fig:kineticTermKN} shows the rotational part of the kinetic term (\ref{KineticKN}) that appears in the integral (\ref{IntegralKN}) for some of the numerical solutions depicted in Figs. \ref{fig:RadialKN2} and \ref{fig:RadialKN3} with parameters displayed in Tables~\ref{tab:spectrum1KN}-\ref{tab:spectrum3KN}, and evaluated at
$\theta = \pi/2$, for simplicity.
We observe that on the horizon this quantity is positive but then becomes negative. This rotational part has been normalized with the square of the field amplitude $\Psi^*\Psi=\phi^2$, which is positive. Furthermore, due to this normalization, the quantity $\mathcal{R}=\Lambda/\phi^2$ does not vanish asymptotically, but it is close to the negative constant
value given by Eq.(\ref{RKNinfty}) $\mathcal{R}=\Lambda/\phi^2
\rightarrow - \left(m\Omega_H + q\Phi_H\right)^2 = - \omega^2$.

Table~\ref{tab:kinKN} shows some numerical values of the quantity $\Lambda/\Psi^*\Psi$ at the horizon and asymptotically $(r \rightarrow \infty)$ corroborating that the numerical results are consistent with the analytical expectations.
\begin{table}[h!]
\centering
\begin{tabular}{|c|c|c|c|}
\hline
\multicolumn{4}{|c|}{Kinetic term (rotational contribution)} \\
\hline
 $m$ & $\mu r_H$ & $\dfrac{\Lambda}{\Psi^*\Psi }\left(r = r_{H}\right)$ & $\dfrac{\Lambda}{\Psi^*\Psi }\left(r \rightarrow \infty \right)$\\
\hline \hline
\multirow{3}{0.2cm}{1} & 0.15 & 43.77526399 & -0.99998181\\ \cline{2-4}
& 0.40 & 4.63855257 &  -0.99516743\\ \cline{2-4}
& 0.57 & 0.75238174 & -0.93941249\\ \hline
\multirow{3}{0.2cm}{2} & 0.50 & 14.36173016 & -0.99668771\\ \cline{2-4}
& 0.95 & 2.400727382 & -0.96461025\\ \cline{2-4}
& 1.17 & 0.67682839 & -0.80623325 \\ \hline
\multirow{3}{0.2cm}{3} & 0.50 & 34.38453473 & -0.99829089\\ \cline{2-4}
& 1.00 & 7.11593895 & -0.98607769\\ \cline{2-4}
& 1.86 & 0.61079886 & -0.69454971\\ \hline
\end{tabular}
\caption{Values of the rotational contribution to the kinetic term at $r = r_{H}$ and $r \rightarrow \infty$ associated with Figure~\ref{fig:kineticTermKN}.}
\label{tab:kinKN}
\end{table}

The fact that the rotational contribution is negative in most of the DOC for the KNBH indicates that the integral (\ref{IntegralKN}) vanishes due to the presence of such a negative quantity, without the need for the field
$\Psi(t, r, \theta, \varphi)$ to vanishes identically, something that does occur in the spherically symmetric scenario \cite{Bekenstein1972,Bekenstein1995,Sudarsky1995,Pena1997,Sudarsky1998}, and which leads to the no-hair theorems in such scenario.
This situation is quite similar to what is found in the background of a Kerr black hole \cite{Garcia2019}.
These results allow us to understand in a simple and heuristic way the existence of nontrivial charged scalar clouds
and the explain why the no-hair theorems cannot be extended for rotating BH's precisely due to
the presence of the rotational contribution $\Lambda$ in the integrand of Eq.(\ref{IntegralKN}) which is not positive
(semi) definite when non-trivial cloud solutions exist.


\section{Extremal scenarios: a new perspective}
\label{Extremalscenario}

In a previous investigation \cite{Garcia2020} we analyzed the existence of (uncharged) clouds in the background
of an {\it exact} extremal Kerr BH. In order to do so, we imposed some {\it superregularity} conditions
at the horizon $r_H^{\rm ext}=M$ on the radial part $R(r)$ of the field $\Psi$. Such conditions consisted
in assuming that $R(r)$ and its derivatives were bounded at the horizon, and contrary to what happens in the
subextremal scenario, those conditions constrained
the separation constants $K_{lm}$ in such a way that {\it a priori} they did not coincide with the values (\ref{KlmKN})
required for the angular part $S_{lm}(\theta)$ of the spheroidal harmonics to be regular on the axis of symmetry
$\theta = \{0, \pi\}$.
These inconsistencies forced us to conclude that regular clouds could not exist around {\it exact} extremal Kerr BH's
unless the pairs of numbers $(l,m)$ satisfied a Diophantine equation of Pell type and only if the numbers of nodes
$n$ were arbitrarily large. Notwithstanding, even in that case, those clouds seemed not to coincide with the
clouds obtained from the subextremal scenario in the limit of extremality $M\rightarrow a$:  both solutions were not connected in
a ``continuous'' way, and as mentioned, the exact extremal ones required the existence of a large number of nodes,
and an additional restriction on the numbers $(l,m)$ (a Pell-Diophantine equation), something that does not occur for
cloud solutions around subextremal Kerr BH's.

If we follow the same strategy for exact extremal KNBH we encounter similar kind of problems, which allow us 
to conclude that assuming superregularity conditions for the radial part $R(r)$ is a too restrictive a condition,
which as we show next, it is not necessary in order to have well behaved clouds.

Now, if the derivatives of
$R(r)$ are not bounded at the horizon, the question is, how to solve the problem, and clearly, how can the corresponding
clouds be {\it regular} at the horizon. The answer to these questions are analyzed in this section by implementing a new method
that leads to regular clouds even if the radial function $R(r)$ has unbounded derivatives at the horizon. Moreover, these cloud solutions
approach the cloud solutions from the subextremal scenario in the limit of extremality, all without the need of
constraining the separation constants and without requiring a large number of nodes, as we will see.

The key aspect to take into account is that first derivatives $R'(r)$ that appear whether in the trace of the energy-momentum
tensor for the scalar field or in the kinetic part $K$ of the action functional of the theory in the form
$g^{ab}(D_a \Psi)^* D_b \Psi=
g^{ab}\nabla_a \Psi^* \nabla_b \Psi$ (when $Q=0$), which 
has an invariant (coordinate independent) meaning, show up
only in the combination $g^{rr} (R'(r))^2$ [times some angular functions; cf. Eq.(\ref{KineticKN})].
In particular, for extremal BH's $g^{rr}$ has a factor $(r-M)^2$, thus $(r-M)^2 R'^2$ can be bounded at the horizon
even if $R'$ blows up there, provided that near the horizon $R'\sim (r-M)^\sigma$ with $\sigma\geq -1$. That is,
provided that near the horizon $R(r) \sim (r-M)^\alpha$ with $\alpha>0$, where $\sigma=\alpha-1$. In this way, not only
the scalar field is bounded at the horizon $r=M$, but the scalar invariants formed from first derivatives of the field
are bounded there as well.

Since in this section we focus only on extremal BH's, we consider the metric (\ref{metricKN}) but taking
$M^2= a^2 + Q^2$, i.e., $r_H^{\rm ext}= M$ which leads to
\begin{equation}
\Delta = (r - M)^2 \;.
\end{equation}

Thus, the complex-valued scalar field $\Psi$ obeys the Klein-Gordon Eq.(\ref{KGequationKN}) in the extremal Kerr
background which is solved again using separation of variables with
the mode expansion like in Eq.(\ref{ScalarFieldKN}), similar to the subextremal scenario.

In order to obtain \textit{bound states} in the extremal scenarios we now assume the following
ansatz factorization for the radial function $R_{nlm}$:
\begin{equation}
  \label{ansatzExt}
R_{nlm}(r) = (r - M)^{\alpha}L_{nlm} (r)\;,
\end{equation}
where the exponent $\alpha$ will be determined from the radial equation for $R_{nlm}(r)$, but it will be required to
be positive (i.e. $\alpha > 0$) such that $R(r_H) = 0$ and $L(r_H)$ is bounded. Otherwise, if $\alpha < 0$,
$R(r_H)$ is unbounded and regular clouds are not possible. Moreover even if $0< \alpha$, the radial derivatives may
still diverge at the horizon if $0< \alpha < 1$. However, this divergence is not physically meaningful as  
the physically meaningful clouds are those with a bounded {\it kinetic term}, 
$K=g^{ab} (D_a\Psi)^* D_b\Psi$ in the DOC, notably at the horizon, in particular, the term $g^{rr} (R')^2$
(where for convenience, we omit the mode labels $(n,m,l)$). More specifically, given (\ref{ansatzExt}),
and for simplicity $Q=0$, the first radial derivative reads,
\begin{equation}
  R'(r) = \alpha(r -M)^{\alpha - 1}L(r) + (r - M)^{\alpha}L'(r)\;,
  \label{Rpextremal}
\end{equation}
which as we emphasized before, it diverges at $r = M$ if $0 < \alpha < 1$. Nevertheless, when considering the term
$g^{rr}R'^2$ that appears in the \textit{kinetic term} $K$, and using the form of the derivative
$R'(r)$ as above, we find
\begin{eqnarray}
  \nonumber g^{rr}R'^2 S^2(\theta)&=& \Big\{
  \frac{\alpha^2}{\rho^2}\left[(r - M)^{\alpha}L(r)\right]^2  \\
\nonumber &+& \frac{2\alpha}{\rho^2}(r - M)^{2\alpha+1}L(r)L'(r) \\ 
 &+& \frac{1}{\rho^2}(r - M)^{2\alpha+2}[L'(r)]^2 \Big\} S^2(\theta)\;,
\label{grrRpsq}
\end{eqnarray}
which is bounded in the DOC for any $\alpha >0$ as far as $L(r)$ and its first derivative are bounded there.
When this is the case, in
particular, at the horizon $r=M$, the term (\ref{grrRpsq}) vanishes there, but it diverges if $\alpha < 0$.

As we will show below, notably, for the extremal Kerr and the extremal KN BH's, we will be looking for
radial solutions of the form (\ref{ansatzExt}) with $0 < \alpha$ and with $L(r)$ and its derivatives bounded 
such that the term (\ref{grrRpsq}) remains bounded in the DOC, despite the fact that $R'$ itself may blow up at the
horizon $r=M$. In this way, the physically meaningful solutions will have $\psi$ and the kinetic term bounded
in the DOC. To achieve this, the angular part for $\psi$, which corresponds to the spheroidal harmonics, also
satisfies suitable regularity conditions, namely, at the axis of symmetry.

\subsection{Extremal Kerr}
\label{subsec:Kerr}
First we analyze solutions for a non-charged field $\Psi$ in the background of a extremal Kerr BH ($|a| = M$; for
concreteness we take $a=M$)  with the metric given in Boyer-Lindquist coordinates by
\begin{eqnarray}\label{metricKerrExt}
\nonumber ds^2 = &-& \left(\frac{\Delta - M^2\sin^2\theta}{\rho^2}\right)dt^2 + \frac{\rho^2}{\Delta}dr^2 + \rho^2d\theta^2 \\
\nonumber &-& \frac{2M\sin^2\theta\left(r^2 + M^2 - \Delta\right)}{\rho^2}dtd\varphi \\
&+& \left(\frac{\left(r^2 + M^2\right)^2 - \Delta M^2\sin^2\theta}{\rho^2}\right)\sin^2\theta d\varphi^2,
\end{eqnarray}
where
\begin{equation}
\rho^2 = r^2 + M^2\cos^2\theta,
\end{equation}
and $M$ is the mass associated with the Kerr BH. In this scenario $q \equiv 0$ and the operator Eq.(\ref{operatorKN}) reduces to the covariant derivative compatible with the spacetime metric:
\begin{equation}
D_a = \nabla_a\;.
\end{equation} 

The radial function $R_{nlm}$ obeys then a simplified version of Eq.~(\ref{Teukolsky-KN}) given by
\begin{eqnarray}\label{radialEKerr}
\nonumber &&\Delta\frac{d}{dr}\left(\Delta\frac{dR_{nlm}}{dr}\right) \\
\nonumber &+&\left[\mathcal{H}_{\rm Kerr}^2+ \left(2mM\omega - K_{lm} - \mu^2\left(r^2 + M^2\right)\right)\Delta\right]\\ 
&\times& R_{nlm} = 0\;,
\end{eqnarray}
where
\begin{equation}
\label{H}
\mathcal{H}_{\rm Kerr} \equiv \left(r^2 + M^2\right)\omega - Mm\;.
\end{equation}
The frequency (\ref{synchronicity}) associated with an extremal Kerr BH and its angular velocity become,
\begin{eqnarray}
  \label{omegaKerr}
\omega &=& m\Omega_H^{\rm ext}\;,  \\
\Omega_H^{\rm ext} &=&\frac{1}{2M}\;.
\end{eqnarray}
When replacing the ansatz (\ref{ansatzExt}) in the radial Eq.(\ref{radialEKerr}) we obtain the following differential equation for $L(r)$ (for brevity we omit the labels $(n,l,m)$ ):
\begin{widetext}
  \begin{equation}
    \label{RadialL}
\left(r - M\right)^{2}L''(r) + 2\left(\alpha + 1\right)(r - M)L'(r)
+ \left[\frac{m^2}{4M^2}\left(r + M\right)^2 + \alpha(\alpha + 1) + m^2 - K_{lm} - \mu^2\left(r^2 + M^2\right)\right]L(r) = 0\;.
\end{equation}
\end{widetext}

\begin{figure}[h]
\begin{center}
\includegraphics[width=0.5\textwidth]{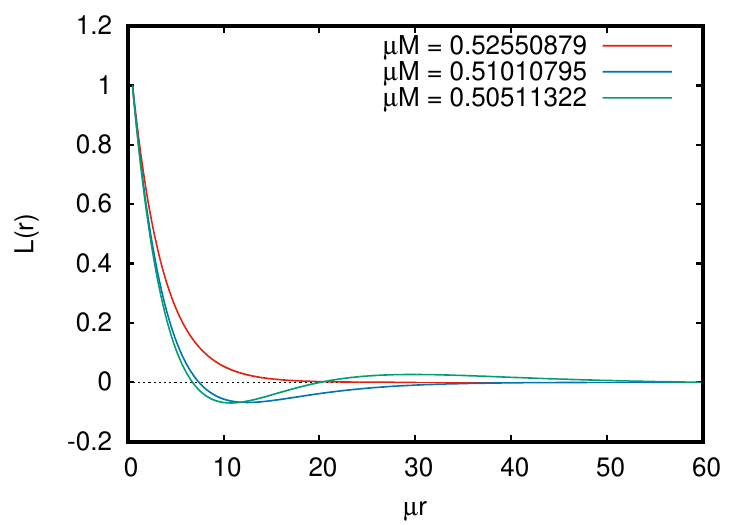}
\includegraphics[width=0.5\textwidth]{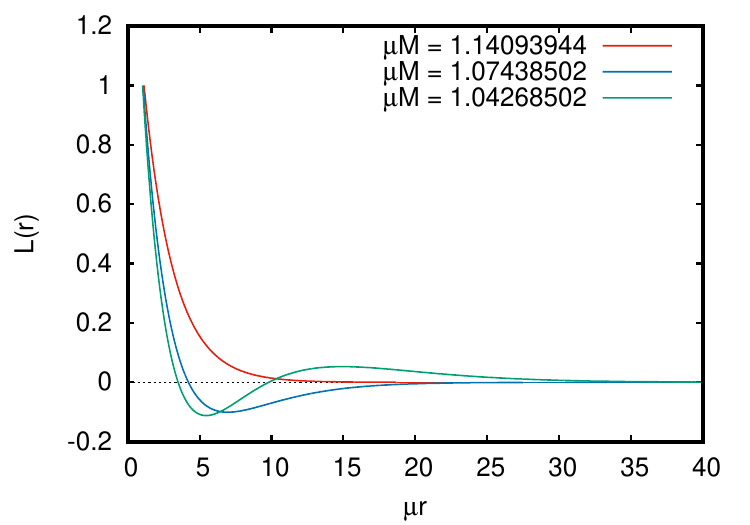}
\caption{(Extremal Kerr scenario) Functions $L(r)$ with $m = l = 1$ (top panel) and $m = l = 2$ (bottom panel), for $n = 0, 1, 2$.
  The functions are bounded and well behaved in the DOC, in particular, at the horizon $r=M$ where $L(M)=1$.} \label{fig:LKerr}
\end{center}
\end{figure}

\begin{figure}[h]
\begin{center}
\includegraphics[width=0.5\textwidth]{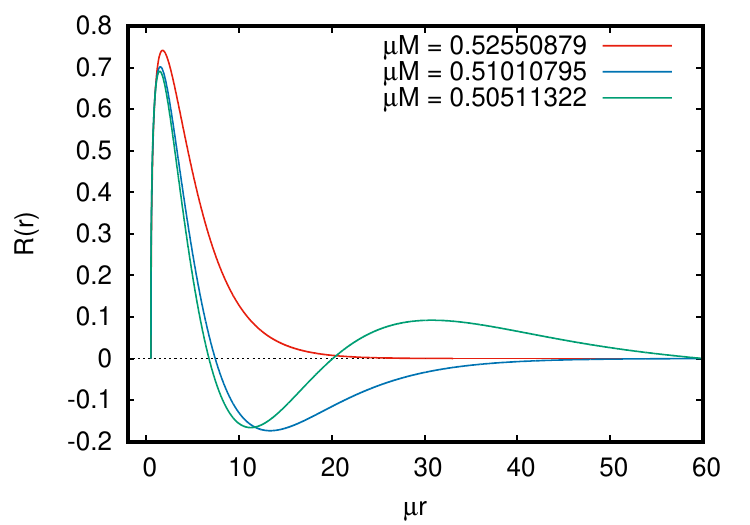}
\includegraphics[width=0.5\textwidth]{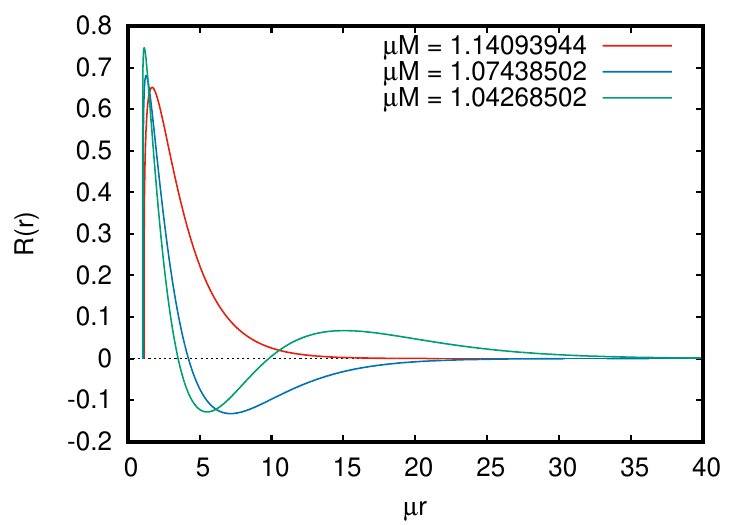}
\caption{Extremal Kerr scenario: (Top panel) Radial Functions ($m = l = 1$) for $n = 0, 1, 2$. The values $\alpha_n$ for each $n$ are $\alpha_0 = 0.38396168$, $\alpha_1 = 0.37307025$ and $\alpha_2 = 0.36957870$, respectively. (Bottom panel) Radial Functions ($m = l = 2$) for $n = 0, 1, 2$ with values $\alpha_n$ given by $\alpha_0 = 0.27115038$, $\alpha_1 = 0.15291680$ and $\alpha_2 = 0.09129800$, respectively. Notice that $R(r)$ vanishes at the horizon $r=M$ and $R'$ blows up there.}\label{fig:RadialKerr}
\end{center}
\end{figure}

Differentiating Eq.~(\ref{RadialL}) once and twice and assuming boundedness of higher derivatives for $L(r)$ at the horizon
we obtain the values (regularity conditions) for $L'(r_H)$ and $L''(r_H)$, which are are given by
\begin{equation}
  \label{RCKerr1}
L'(r_H) = \frac{\left(2\mu^2M^2 - m^2\right)}{2(\alpha + 1)M}L(r_H)\;,
\end{equation}
and
\begin{widetext}
\begin{equation}\label{RCKerr2}
L''(r_H) = \frac{1}{(4\alpha + 6)}\left[\left(4\mu^2M - \frac{2m^2}{M}  \right)L'(r_H) + \left(2\mu^2 - \frac{m^2}{2M^2}\right)L(r_H)\right].
\end{equation}
\end{widetext}
Moreover, given those regularity conditions, the boundedness of $L(r)$ at the horizon and assuming $L(r_H)\neq 0$ we find,
when evaluating (\ref{RadialL}) at $r=M$, that the exponent $\alpha$ satisfies the following simple quadratic algebraic equation
\begin{equation}
  \label{alphaKerr}
\alpha^2 + \alpha + 2m^2 - 2\mu^2M^2 - K_{lm} = 0\;,
\end{equation}
which has the following solutions:
\begin{equation}
  \label{alphaKerr2}
\alpha_{\pm} = \frac{-1 \pm \sqrt{1 + 4\left[K_{lm} + 2\left(\mu^2M^2 - m^2\right)\right]}}{2}\;.
\end{equation}
We choose the sign $(+)$ for avoiding a negative definite $\alpha$ and thus to consider only potentially bounded radial solutions at the horizon. This will be possible provided the positive square root is larger than $-1$. As we show below,
this depends on the values of the BH parameter $\mu M$ (i.e. $M$ in units of $1/\mu$)
and the numbers $(n,l,m)$, notably, the value $m$. Figure~\ref{fig:LKerr} shows solutions for the radial function
$L(r)$ (taking $L(r_H^{\rm ext})=1$ for simplicity) associated with an extremal Kerr BH with ``quantum numbers'' $m = l = 1$, $m = l = 2$ and $n = 0, 1, 2$.
Figure~\ref{fig:RadialKerr} depicts the corresponding radial part $R(r)$ (\ref{ansatzExt}). Each solution has
an exponent $\alpha_+$ that satisfies the condition $0<\alpha_+$ which leads to physically
meaningful regular clouds with a bounded kinetic term, despite the fact that $R'(M)$ diverges given that
$0<\alpha_+ <1$ (see the figure's caption).

\begin{figure}[h]
\begin{center}
\includegraphics[width=0.5\textwidth]{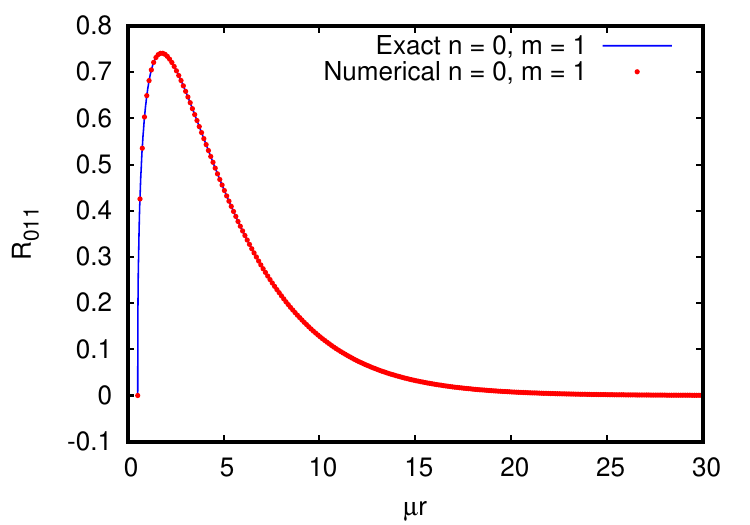}
\includegraphics[width=0.5\textwidth]{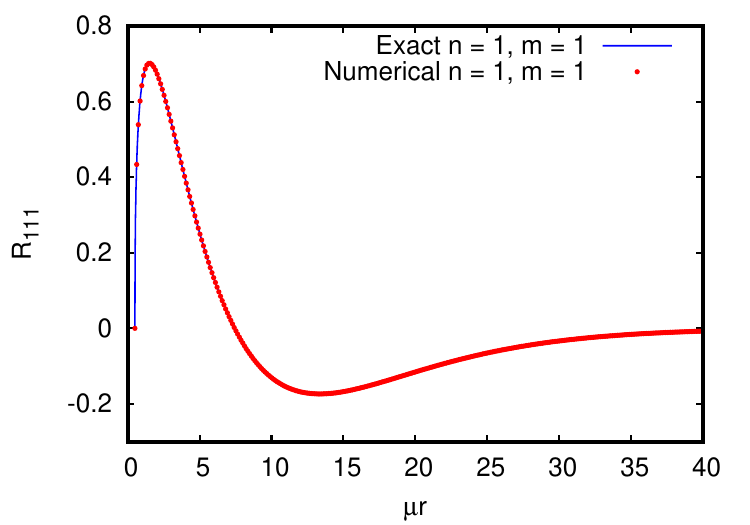}
\caption{Radial solutions $R_{nlm}$ for the extremal Kerr scenario with principal numbers $n = 0, 1$ and $m = l = 1$. The red dots correspond to the numerical solutions obtained by solving Eq. (\ref{Teukolsky-KN}) and the blue continuous line corresponds to the exact solution computed by Hod (Eq.(17) in \cite{Hod2012}). The agreement between both solutions is excellent.}\label{fig:RadialHodKerr}
\end{center}
\end{figure}

\begin{figure}[h]
\begin{center}
\includegraphics[width=0.5\textwidth]{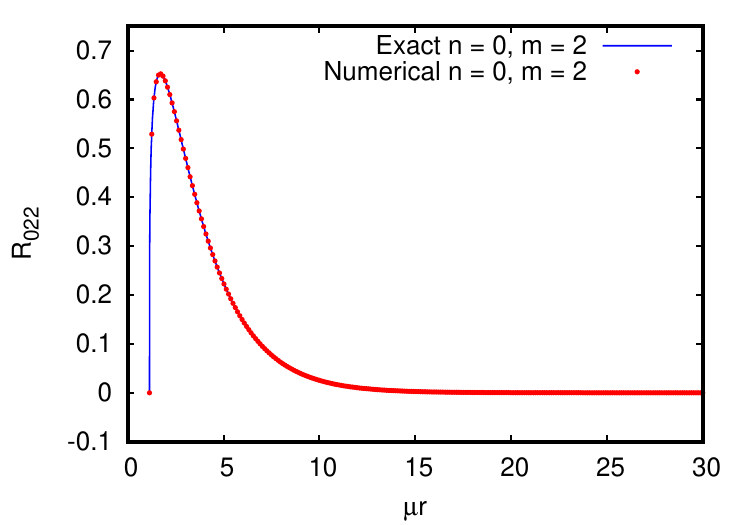}
\includegraphics[width=0.5\textwidth]{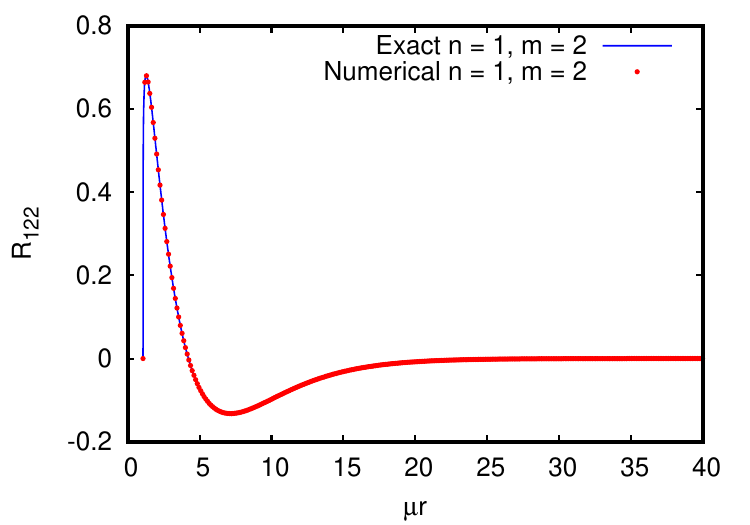}
\caption{Same as Fig.\ref{fig:RadialHodKerr} but taking $n = 0, 1$ and $m = l = 2$.}
\label{fig:RadialHodKerr2}
\end{center}
\end{figure}

A remarkable fact is that Hod \cite{Hod2012} had found
exact solutions to the radial Eq.(\ref{Teukolsky-KN}) associated with this scenario (uncharged clouds in the background
of an extremal Kerr BH, $Q=0$, $a=M$). These solutions are,
\begin{equation}
  \label{HodKerr}
R(z) = Az^{-\frac{1}{2} + \beta}e^{-\frac{1}{2}z}L_{n}^{(2\beta)}(z)\;,
\end{equation}
where
\begin{equation}
z \equiv 2\sqrt{\mu^2 - \frac{m^2}{4M^2}}(r - M)\;,
\end{equation}
and
\begin{equation}
  \label{Hodbeta}
\beta^2 \equiv K_{lm} + \frac{1}{4} - 2m^2 + 2\mu^2M^2\;,
\end{equation}
$L_{n}^{(2\beta)}(z)$ are the generalized Laguerre polynomials and $A$ is a normalization constant. By comparing
Eqs. (\ref{alphaKerr2}) and (\ref{Hodbeta}) we find that the
exponent $\alpha_+$ and Hod's $\beta$ maintain the following relation:
\begin{equation}
\beta = \alpha_{+} + \frac{1}{2}\;.
\end{equation}
Figure~\ref{fig:RadialHodKerr} compares our numerical solution for the radial function $R_{nlm}$ using $r_H = M = a$ with that obtained by Hod analytically taking $l = m = 1$ and different values of $n$. Figure~\ref{fig:RadialHodKerr2} shows the comparison between both solutions for $l = m = 2$. The numerical and the exact solutions show an excellent agreement,
proving that our approach is robust.

\begin{figure}[h]
\begin{center}
   \includegraphics[width=0.5\textwidth]{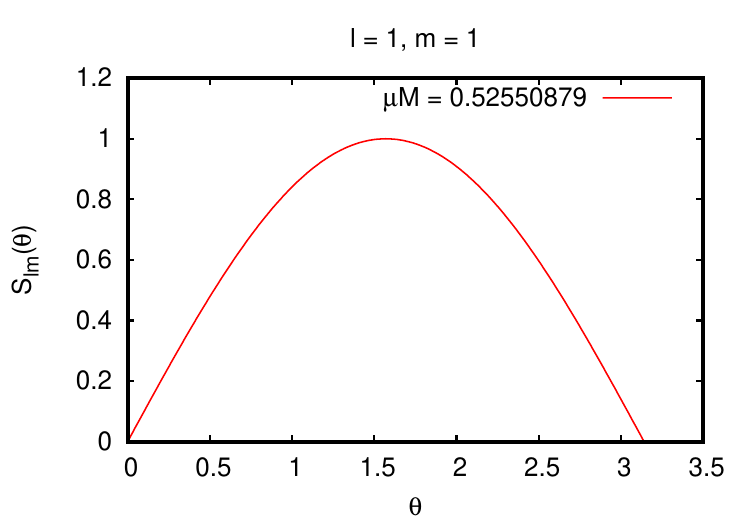}
   \includegraphics[width=0.5\textwidth]{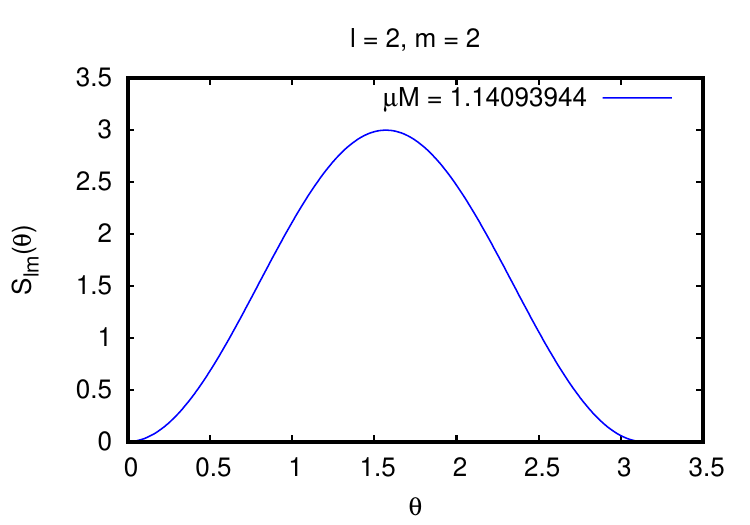}
   \caption{Spheroidal harmonics $S_{lm}(\theta)$ associated with the angular part of the field $\Psi$
     for $l = m = 1$ (top panel) and $l = m = 2$ (bottom panel), respectively. (Extremal Kerr scenario).}\label{fig:AngularPartKerr}
\end{center}
\end{figure}

Figure~\ref{fig:AngularPartKerr} shows the angular functions $S_{lm}(\theta)$ associated with the values
$\mu M = 0.52550879$ ($m = l = 1$) and $\mu M = 1.14093944$ ($m = l = 2$). As we can appreciate, the functions
are perfectly regular on the axis of symmetry at $\theta = \{0, \pi\}$. In this way, we have succeeded
in finding numerically neutral scalar clouds around {\it exact} extremal Kerr BH's which are perfectly
regular at the horizon and on the axis of symmetry. Namely, given the values obtained for the exponent $\alpha$
in each solution, the invariant (coordinate independent) scalar $g^{ab}\nabla_a\Psi^* \nabla_b \Psi$ turns to be bounded in the DOC, notably,
at the horizon even if the radial derivative $R'(r)$ is unbounded there. 

\subsection{Extremal Reissner-Nordstrom} 
\subsection*{Case $q^2 < \mu^2$}
\label{subsec:RN}
Next we consider a charged field $\Psi$ coupled with an extremal RNBH with the metric
\begin{eqnarray}\label{metricRN}
\nonumber ds^2 = &-&\frac{\left(r - M\right)^2}{r^2}dt^2 + \frac{r^2}{\left(r - M\right)^2}dr^2 \\
&+& r^2d\theta^2 + r^2\sin^2\theta d\varphi^2,
\end{eqnarray}
where $M=|Q|$ is the mass associated with the extremal RNBH. For this case we consider the
derivative operator (\ref{operatorKN}) with
\begin{equation}
  A_a = -\frac{Q}{r} dt_a\;,  
\end{equation}
and the scalar-field potential (\ref{PotentialBH}) for a massive free field, like in the Kerr BH of Sec.~\ref{subsec:Kerr}.
The radial Eq. (\ref{Teukolsky-KN}) takes the form 
\begin{equation}\label{radialRN}
\Delta\frac{d}{dr}\left(\Delta\frac{dR_{nlm}}{dr}\right) + \left[\mathcal{H}_{\rm RN}^2 - \left(K_{lm} + \mu^2r^2 \right)\Delta\right]R_{nlm} = 0  
\end{equation}
with
\begin{equation}\label{HRN}
\mathcal{H}_{\rm RN} \equiv \omega r^2 - qMr\;,
\end{equation}
and
\begin{equation}
K_{lm} = l(l + 1)\;.
\end{equation}

In this case, the boson-field frequency (\ref{synchronicity}) associated with an extremal RNBH is
\begin{equation}\label{omegaRN}
\omega = q\;.  
\end{equation} 

The radial equation for the function $L(r)$ in this scenario is 
\begin{eqnarray}\label{RadialERN}
&&\nonumber (r - M)^2L'' + 2(\alpha + 1)(r - M)L' \\
&+& \left[\alpha(\alpha + 1) + \left(q^2 - \mu^2\right)r^2 - l(l + 1)\right]L  = 0\,, 
\end{eqnarray}
subject to the following regularity conditions:
\begin{equation}
  \label{RCRN1}
L'(r_H) = \frac{\left(\mu^2 - q^2\right)M}{\alpha + 1}L(r_H)\;,
\end{equation}
and
\begin{equation}
  \label{RCRN2}
L''(r_H) = \frac{\left(\mu^2 - q^2\right)}{2\alpha + 3}\left[2ML'(r_H) + L(r_H)\right]\;,
\end{equation}
with $\alpha$ satisfying the following quadratic equation
\begin{equation}\label{alphaRN}
\alpha^2 + \alpha + \left(q^2 - \mu^2\right)M^2 - l(l + 1) = 0\;,
\end{equation}
with solutions
\begin{equation}\label{alphaRN2}
\alpha_{\pm} = \frac{-1 \pm \sqrt{1 + 4\left[l(l + 1) + \left(\mu^2 - q^2\right)M^2\right]}}{2}\;.
\end{equation}
As before, we take only the sign $(+)$ in order for $\alpha$ to be positive and the radial function $R(r)$
to be bounded at $r=M$.

Using the value for $\alpha_{+}$ in Eq.~(\ref{alphaRN2}) the radial Eq.~(\ref{RadialERN}) for $L(r)$ reads as follows:
\begin{equation}\label{RadialERN2}
(r - M)L'' + 2(\alpha_{+} + 1)L' + \left(q^2 - \mu^2\right)(r + M)L = 0\;. 
\end{equation}

When we solve numerically Eq.(\ref{RadialERN2}) using the regularity conditions (\ref{RCRN1}) and (\ref{RCRN2})
we find that the only acceptable solution is 
\begin{equation}
L(r) = \text{const}\;.
\end{equation} 
Since we demand that the boson field vanishes asymptotically we conclude that the function $L(r)$ is the trivial solution
$L(r) \equiv 0$, so that the radial functions $R_{nlm}(r)$ is well behaved asymptotically. Thus, under such circumstances 
$R_{nlm}(r)$ also vanishes everywhere and the only solution to the radial Eq.(\ref{RadialERN}) in this scenario is
\begin{equation}
R_{nlm}(r) \equiv 0\;.  
\end{equation}  
These numerical results corroborate that it is not possible to find charged scalar clouds around an extremal Reissner-Nordstrom black hole, as it was concluded previously in Ref.\cite{Garcia2021}. There exists, however, a way to avoid this conclusion,
but it requires a scalar-field potential that includes a self-interaction term \cite{Garcia2021,Hong2020,Herdeiro2020,Brihaye2021}. Under such conditions it is possible to find numerically charged clouds, known in the literature as Q-clouds,
which as we just concluded, are absent when the field is only a massive but free.

\subsection*{Double extremal case $q^2 = \mu^2$}

Considering the case $q^2 = \mu^2$ analyzed in \cite{DegolladoHerdeiro2013,Garcia2021} Eq.~(\ref{RadialERN2}) reads
\begin{equation}\label{RadialDextRN1}
(r - M)L'' + 2(\alpha_+ + 1)L' = 0\,.
\end{equation}
However, we see that in this double extremal scenario Eq.~(\ref{alphaRN2}) leads to the following possible values for $\alpha_+$ 
\begin{equation}\label{RadialDextRN2}
\alpha_+ = l\;.
\end{equation}
So Eq.~(\ref{RadialDextRN1}) takes the form
\begin{equation}\label{RadialDextRN3}
(r - M)L'' + 2(l + 1)L' = 0\;,
\end{equation}
whose solution is of the form
\begin{equation}
  L(r) = \frac{const.}{(2l + 1)\left(r - M\right)^{2l + 1}}\;,
  \label{LRNdext}
\end{equation}
so the radial function $R(r) = (r - M)^{\alpha}L(r)$ is
\begin{equation}
R(r) = \frac{const.}{(2l + 1)\left(r - M\right)^{l + 1}}\;.
\end{equation}
We can see that these solutions are not regular at horizon $r = M$ for any $l \geqslant 0$, and therefore, the
only possibility is again $R(r)\equiv 0$, a result that is in a agreement with the results reported in \cite{Garcia2021}.
In fact, from (\ref{RCRN1}) and (\ref{RCRN2}) we already notice that when
  $q^2 = \mu^2$, then $L'(r_H)\equiv 0 \equiv L''(r_H) $, and thus,
  the solution (\ref{LRNdext}) requires the constant to be identically zero, leading also to $R(r)\equiv 0$.

\subsection{Extremal Kerr-Newman}
\label{subsec:KerrNewman}

We now consider a charged field $\Psi$ coupled to an extremal KNBH with the metric (\ref{metricKN}) but taking $\Delta = (r - M)^2$ and $M^2 = a^2 + Q^2$. We assume again the scalar-field potential for a free but massive field (\ref{PotentialBH}).

The equation associated with the radial part of the scalar field is provided by Eq.~(\ref{Teukolsky-KN}). The
frequency for an extremal KNBH is the the same as in (\ref{synchronicity}) except that the BH's angular velocity is  
\begin{equation}
\Omega_H \equiv \frac{a}{M^2 + a^2}\;,
\end{equation}
and the electric potential at the horizon is
\begin{equation}
\Phi_H \equiv \frac{QM}{M^2 + a^2}\;.
\end{equation}

As before, the integer $m$ is the azimutal harmonic index and $q$ is the charge coupling constant of the scalar field. The radial equation for the function $L(r)$ now satisfies the following differential equation:
\begin{widetext}
  \begin{equation}
    \label{RadialEKN}
(r - M)^2L'' + 2(\alpha + 1)(r - M)L' + \left\lbrace\alpha(\alpha + 1) + \left[\frac{am(r + M) + qQ\left(Mr - a^2\right)}{M^2 + a^2}\right]^2 + 2ma\omega - K_{lm} - \mu^2\left(r^2 + a^2\right)\right\rbrace L = 0\;.
\end{equation}
\end{widetext}
The regularity conditions for the first and second derivatives of $L(r)$ at the horizon are 
\begin{widetext}
  \begin{equation}
    \label{RCKN1Ext}
L'(r_H) = \frac{\mu^2M\left(M^2 + a^2\right)^2 - (am + qQM)(2amM + qQ^3)}{(\alpha + 1)\left(M^2 + a^2\right)^2}L(r_H)\;,
\end{equation}
\end{widetext}
\begin{widetext}
  \begin{eqnarray}
    \label{RCKN2Ext}
\nonumber L''(r_H) = && \frac{1}{(2\alpha + 3)\left(M^2 + a^2\right)^2}\left\lbrace \left[\mu^2\left(M^2 + a^2\right)^2 - (am + qQM)^2\right]L(r_H)\right.\\
&+& 2\left.\left[\mu^2M\left(M^2 + a^2\right)^2 - \left(2maM + qQ^3\right)\left(am + qQM\right)\right]L'(r_H)\right\rbrace \,,
\end{eqnarray}
\end{widetext}
respectively, with $\alpha$ satisfying the following equation 
\begin{widetext}
  \begin{equation}
    \label{alphaKN}
\alpha^2 + \alpha + \left[\frac{2amM + qQ^3}{M^2 + a^2}\right]^2 + 2ma\omega - K_{lm} - \mu^2\left(M^2 + a^2\right) = 0\;,
\end{equation}
\end{widetext}
whose solutions are 
\begin{widetext}
  \begin{equation}
    \label{alphaKN2}
\alpha_{\pm} = \frac{-1 \pm \sqrt{1 + 4\left\lbrace K_{lm}  + \mu^2\left(M^2 + a^2\right) - \left[2ma\omega\left(M^2 + a^2\right)^2 + \left(2amM + qQ^3\right)^2\right]/\left(M^2 + a^2\right)^2\right\rbrace}}{2}\;.
\end{equation}
\end{widetext}

Again, we take the sign $(+)$ for $\alpha$ to be positive. We appreciate that when taking $Q = 0$ but $a \neq 0$ in the regularity conditions (\ref{RCKN1Ext}) and (\ref{RCKN2Ext}) and also in Eq.~(\ref{alphaKN2}) for $\alpha$, they reduce to Eqs.~(\ref{RCKerr1}), (\ref{RCKerr2}) and (\ref{alphaKerr2}), respectively, which are associated with the extremal Kerr BH. On the other hand, when we take $a = 0$ and $Q \neq 0$ in those equations, we recover Eqs.(\ref{RCRN1}), (\ref{RCRN2}) and (\ref{alphaRN2}), respectively, corresponding to the extremal RNBH.

\begin{figure}[h]
\begin{center}
\includegraphics[width=0.5\textwidth]{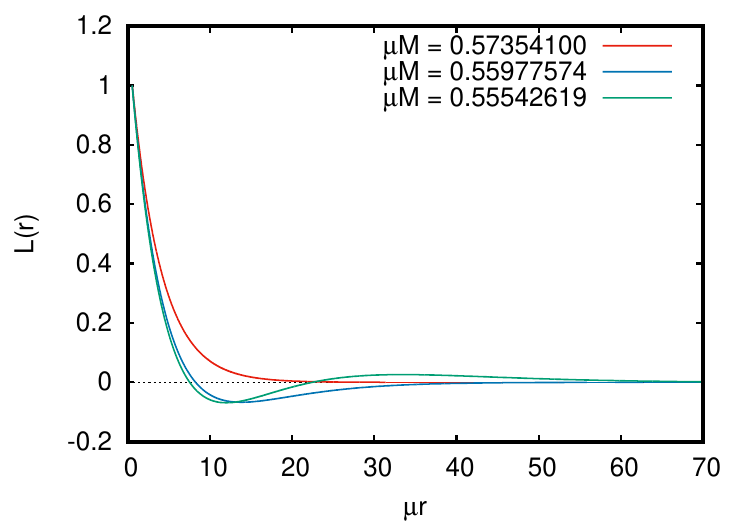}
\includegraphics[width=0.5\textwidth]{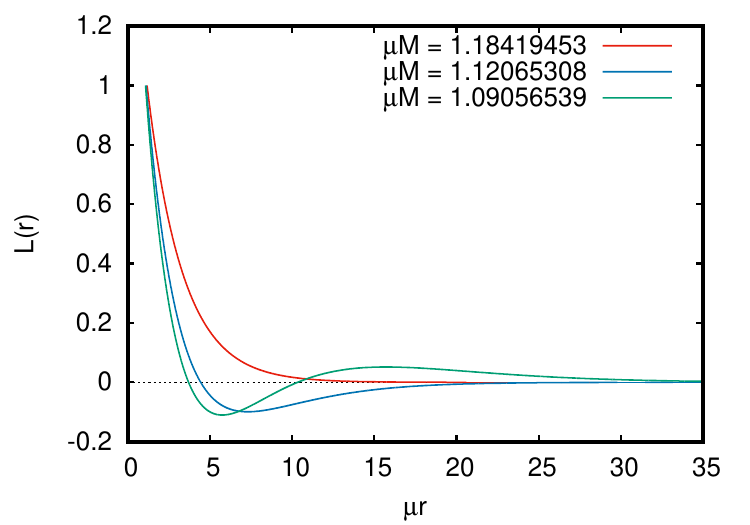}
\caption{Extremal Kerr-Newman scenario $q/\mu = 1$ and $\mu Q = 0.1$: Functions $L(r)$ with $m = l = 1$ (top panel) and $m = l = 2$ (bottom panel), for number
  of nodes $n = 0, 1, 2$. Like in the extremal Kerr scenario, these functions are bounded and well behaved in the DOC, in particular, at the horizon $r=M$ where $L(M)=1$.} \label{fig:LKerrNewman}
\end{center}
\end{figure}

Figure~\ref{fig:LKerrNewman} depicts the radial function $L(r)$ associated with an extremal KNBH
for $m = l = 1$ and $m = l = 2$, with number of nodes $n = 0, 1, 2$. 

\begin{figure}[h]
\begin{center}
\includegraphics[width=0.5\textwidth]{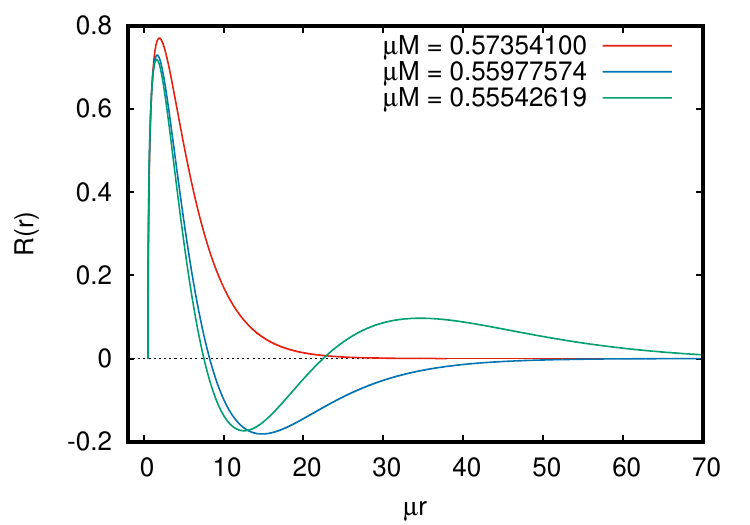}
\includegraphics[width=0.5\textwidth]{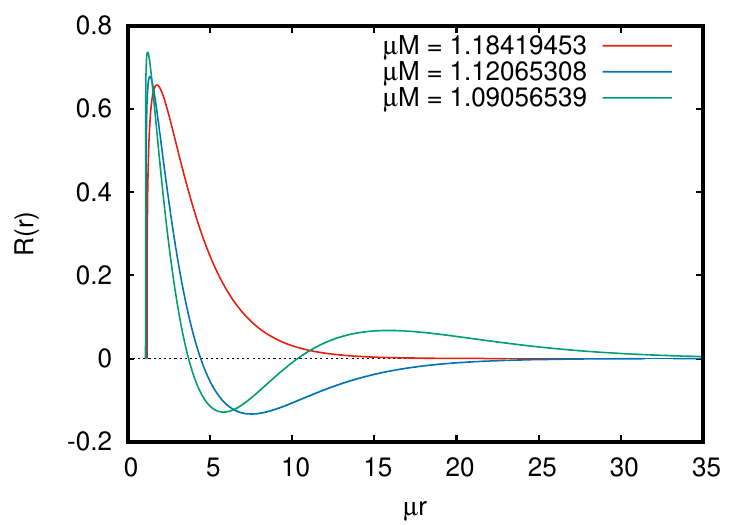}
\caption{Extremal Kerr-Newman scenario $q/\mu = 1$ and $\mu Q = 0.1$: (Top panel) Radial Functions $R$ with $m = l = 1$, for $n = 0, 1, 2$. The values of
  the exponent $\alpha_n$ for each $n$ are $\alpha_0 = 0.38914753$ $(\mu a = 0.56475594)$, $\alpha_1 = 0.37873576$ $(\mu a = 0.55077117)$ and $\alpha_2 = 0.37547687$ $(\mu a = 0.54634993)$, respectively. (Bottom panel) Radial Functions with $m = l = 2$, for $n = 0, 1, 2$. In this case $\alpha_0 = 0.27624214$ $(\mu a = 1.17996470)$, $\alpha_1 = 0.16035789$ $(\mu a = 1.11618248)$ and $\alpha_2 = 0.10040099$ $(\mu a = 1.08597093)$, respectively. Like in the extremal Kerr scenario, notice that $R(r)$ vanishes at the horizon $r=M$ and $R'$ blows up there.}\label{fig:RadialKN}
\end{center}
\end{figure}

Figure~\ref{fig:RadialKN} shows the complete radial part $R(r)$ (\ref{ansatzExt}) 
of the scalar field $\Psi$ around an extremal KNBH corresponding the functions $L(r)$ of Fig.~\ref{fig:LKerrNewman}
for the integers $m = l = 1$ and $m = l = 2$,  and nodes $n = 0, 1, 2$. These solutions have $0 <\alpha_+ <1$
(see the caption) and thus leads to regular clouds with a bounded kinetic term even if $R'$ blows up at
the extremal horizon $r=M$.

Like in the extremal Kerr BH, Hod \cite{Hod2015} also found exact solutions for clouds around extremal KNBH. The solutions to the radial Eq.~(\ref{Teukolsky-KN}) reported by Hod are,
\begin{equation}
  \label{HodKerrNewman}
R(\chi) = A\chi^{-\frac{1}{2} + \beta}e^{-\epsilon\chi}L_{n}^{(2\beta)}(2\epsilon\chi),
\end{equation}
where
\begin{equation}
\chi \equiv \frac{r - M}{M},
\end{equation}
\begin{equation}
\epsilon \equiv M\sqrt{\mu^2 - \omega^2} \,,
\end{equation}
\begin{equation}
\beta^2 \equiv K_{lm} + \frac{1}{4} - 2am\omega - (2M\omega - qQ)^2 + \mu^2\left(M^2 + a^2\right)\,,
\end{equation}
$L_{n}^{(2\beta)}(2\epsilon\chi)$ are the generalized Laguerre polynomials and $A$ is a normalization constant. The
exponent $\alpha$ and Hod's $\beta$ maintain the same relationship as in the extremal Kerr, 
\begin{equation}
\beta = \alpha_{+} + \frac{1}{2}\;,
\end{equation}
except that the $\alpha_{+}$ (\ref{alphaKN2}) includes the electric charges $Q$ and $q$, and it also includes the separation constants $K_{lm}$ corresponding to the charged extremal scenario.

\begin{figure}[h]
\begin{center}
\includegraphics[width=0.5\textwidth]{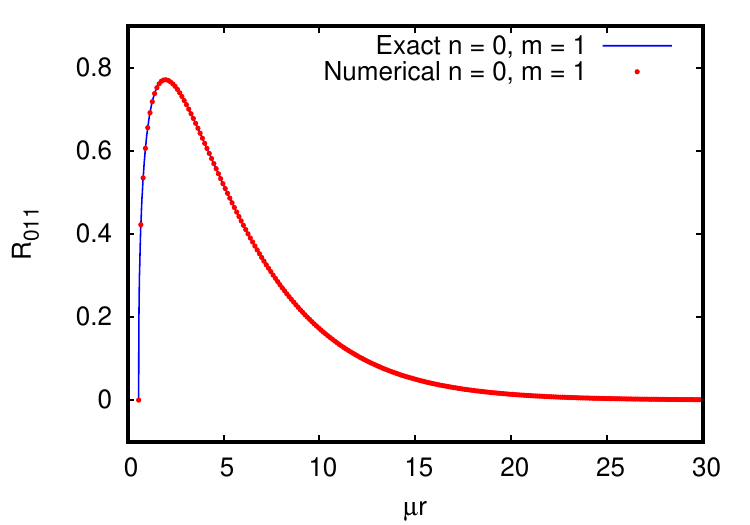}
\includegraphics[width=0.5\textwidth]{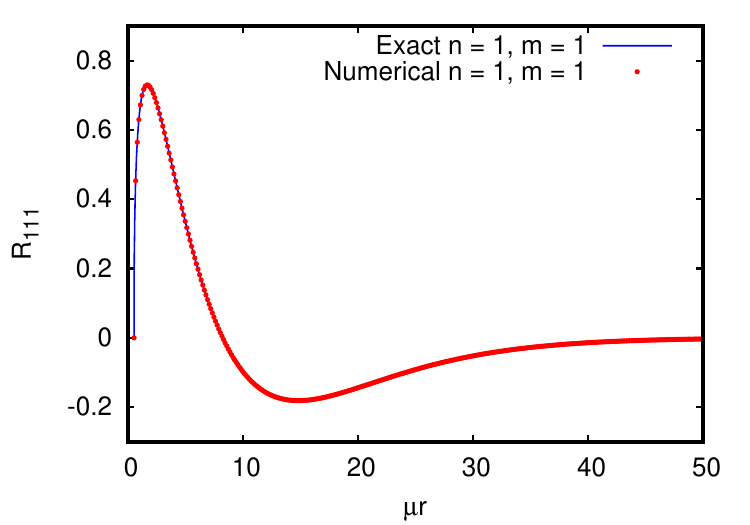}
\caption{Radial solutions $R_{nlm}$ for an extremal KNBH ($q/\mu = 1$ and $\mu Q = 0.1$)
  with principal numbers $n = 0, 1$ and $m = l = 1$. The red dots correspond to the numerical solutions obtained by solving the Eq.(\ref{Teukolsky-KN}) and the blue continuous line
  shows the exact solution given by Hod (Eq.(49) in \cite{Hod2015}). 
Notice the excellent agreement between the analytic and the numerical solutions.}\label{fig:RadialKNHod}
\end{center}
\end{figure}

\begin{figure}[h]
\begin{center}
\includegraphics[width=0.5\textwidth]{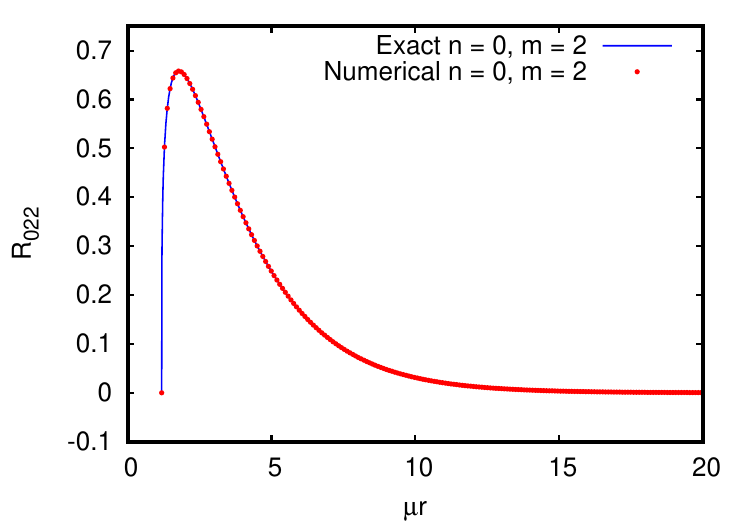}
\includegraphics[width=0.5\textwidth]{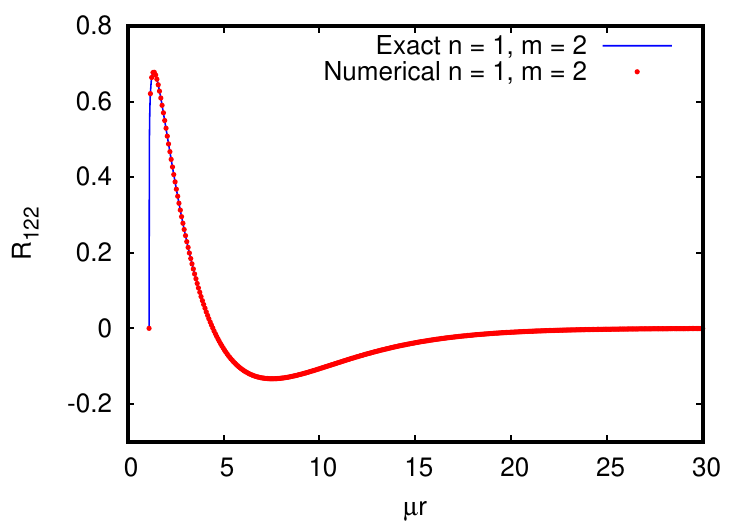}
\caption{Radial solutions $R_{nlm}$ for the extremal KNBH ($q/\mu = 1$ and $\mu Q = 0.1$) with principal numbers $n = 0, 1$ and $m = l = 2$. The red dots correspond to the numerical solutions obtained from solving the equation (\ref{Teukolsky-KN}) and the blue continuous line corresponds to the exact solution given by Hod (Eq.(49) in \cite{Hod2015}).
The agreement between the analytic and the numerical solutions is excellent.}\label{fig:RadialKNHod2}
\end{center}
\end{figure}

Fig.~\ref{fig:RadialKNHod} compares our numerical solution for the radial function $R_{nlm}$
(with $r_H^{\rm ext} = M$) with
Hod's (\ref{HodKerrNewman}) taking $l = m = 1$ and different values for $n$. Figure~\ref{fig:RadialKNHod2}
depicts something similar but taking $l = m = 2$.

\begin{figure}[h!]
   \includegraphics[width=0.5\textwidth]{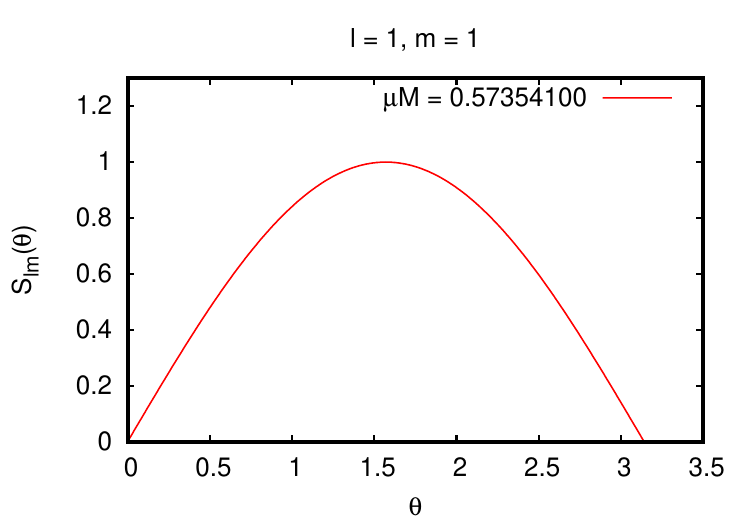}
    \includegraphics[width=0.5\textwidth]{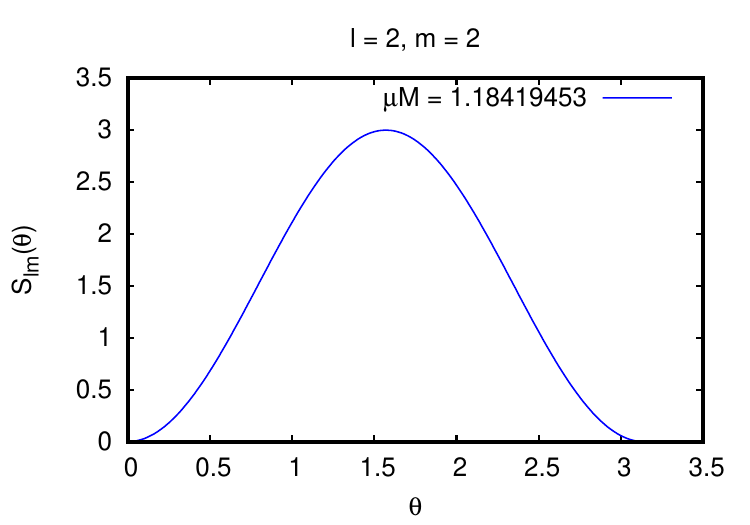}
\caption{Spheroidal harmonics $S_{lm}(\theta)$ associated with the angular part of the charged scalar field $\Psi$ corresponding to take $l = m = 1$ (top panel) and $l = m = 2$ (bottom panel) respectively. (Extremal Kerr-Newman scenario).}\label{fig:AngularPartKN}
\end{figure}

Figure~\ref{fig:AngularPartKN} shows the angular functions $S_{lm}(\theta)$ associated with the charged scalar field $\Psi$ for the values $\mu M = 0.57354100$ ($m = l = 1$) and $\mu M = 1.18419453$ ($m = l = 2$). From the figure one can appreciate
the regularity of those functions on the axis of symmetry. In this way we have obtained a complete regular
solution (namely, regular at the horizon and on the axis of symmetry) for the charged scalar field $\Psi$ around an extremal
KNBH, which in addition leads to a well behaved kinetic term $g^{ab}(D_a \Psi)^* D_b \Psi$ given by Eq.(\ref{KineticKN})
despite the fact that its radial derivative is unbounded at the BH's horizon\footnote{The unboundedness of the radial derivatives of $R(r)$ at the horizon, together that $R(r)$ vanishes there,
are features that can be appreciated also from the 
solutions (\ref{HodKerr}) and (\ref{HodKerrNewman}) found by Hod \cite{Hod2012,Hod2015}
for clouds around extremal Kerr and KN BH's, respectively, and more vividly from Figs.~\ref{fig:RadialHodKerr},
\ref{fig:RadialHodKerr2}, \ref{fig:RadialKNHod} and \ref{fig:RadialKNHod2}.}.

In summary, by using the decomposition for the radial function with the ansatz (\ref{ansatzExt}) it is possible to obtain
both neutral and electrically charged regular scalar cloud solutions around (exact) extremal Kerr and extremal KNBH, respectively.

To conclude this section, a final analysis is in order. Let us remind that in the subextremal scenarios the value of
the radial function at the horizon $R(r_H)$ is a free parameter that for simplicity and convenience
we take it as $R(r_H)=1$. However, if we want that the radial solutions approach as continuously as possible to the corresponding
solutions of extremal BH's we need to modify $R(r_H)$ such that as $r_H\rightarrow r_H^{\rm ext}=M$,
$R(r_H)\rightarrow 0$, since in the extremal scenarios by construction the radial functions
vanish $R(r_H^{\rm ext})\equiv 0$. To achieve this, we propose the following modification for the
value $R(r_H)$:
\begin{equation}
  \label{NewradialSub}
R(r_H) = C\times (r_H - M)^\alpha\;,
\end{equation}
where $0 < \alpha < 1$ is the same exponent as computed for the respective radial solutions for extremal Kerr and KNBH
(i.e. using the same numbers $(l,m,n)$) that we want to compare with. Here $C$ is a constant that we take $C = 1$.
Moreover, taking into account the regularity condition Eq.(\ref{RCKN1}) we observe that
in the extremal limit ($r_H \rightarrow M$) and with the modification (\ref{NewradialSub}) we will have 
\begin{equation}
R(r_H) \rightarrow 0\;,\;\;\;\text{and}\;\;\; R'(r_H) \rightarrow \infty\;,
\end{equation}
which, as a bonus, allows us to avoid a trivial solution that we would obtain if the exponent was fixed
with a value $\alpha > 1$ since then $R'(r_H)\equiv 0$. In this way, the regularity conditions are respected but
are changed suitably for each cloud computed around subextremal BH's (Kerr and Kerr-Newman). In particular,
the value $R(r_H)$ is not fixed anymore to $R(r_H)=1$, but changes with $r_H$ and $M$ according to (\ref{NewradialSub}).

Using this improved method we obtain a ``smooth'' match between the radial solutions associated with the subextremal and extremal
scenarios in the limit where the former approach the latter, something, that as emphasized in the Introduction,
did not happen with the treatment carried out by us in \cite{Garcia2020}.  

\begin{figure}[h!]
   \includegraphics[width=0.5\textwidth]{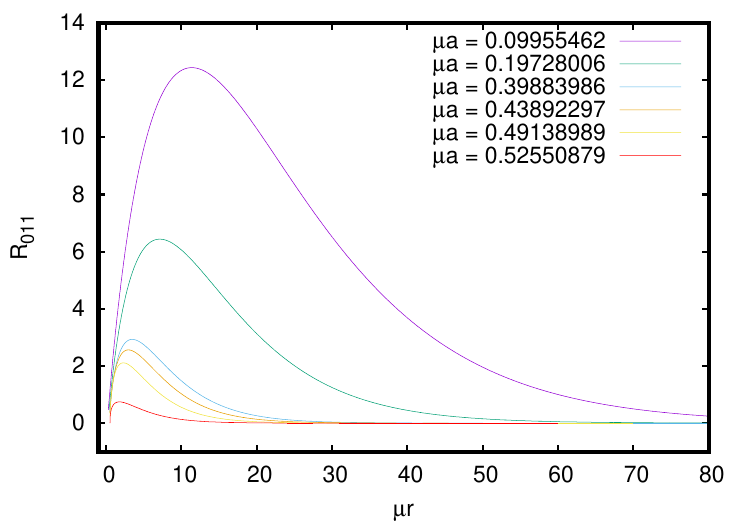}
    \includegraphics[width=0.5\textwidth]{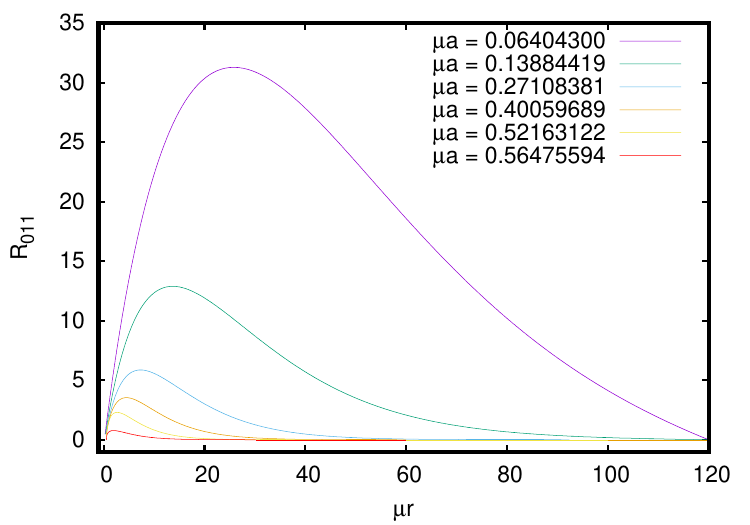}
    \caption{Radial functions $R_{nlm}$ associated with a scalar field $\Psi$ around a subextremal Kerr BH (top panel) and
      KNBH (bottom panel) with $n = 0$, $m = 1$, $l = 1$ for different values of the BH angular momentum
      $a$ that converge to the extremal values $r_H^{\rm ext, Kerr}=M=a\approx 0.525/\mu$ (top panel) and
      $r_H^{\rm ext, KN}= M = \sqrt{a^2+ Q^2}\approx 0.564/\mu$ (bottom panel), with $Q = 0.1/\mu$.
      Here $R(r_H) = (r_H - M)^\alpha$. Notice that the slope as well
      as the value $R(r_H)$ change as the extremal cases are reached. For reference,
      the corresponding radial solutions around the exact extremal BH's (red curves)
      are included in both panels.}\label{fig:RadialNewFunction}
\end{figure}

Figure~\ref{fig:RadialNewFunction} shows a sample of radial cloud solutions $R_{nlm}$ that are recomputed for the
subextremal Kerr BH (top panel) and the subextremal KNBH (bottom panel) with the new value  $R(r_H)$ fixed according to the
prescription (\ref{NewradialSub}) where $\alpha$ is taken, respectively, as $\alpha_+$
of Eq.(\ref{alphaKerr2}) and $\alpha_+$ of Eq.(\ref{alphaKN2}). In each panel, the radial function is approaching
smoothly to the corresponding solution for the exact extremal case
(and the maximum amplitude decreases as a consequence of this) as $r_H\rightarrow M$. The latter are depicted in both panels by the red line solutions. 

\begin{figure}[h!]
   \includegraphics[width=0.5\textwidth]{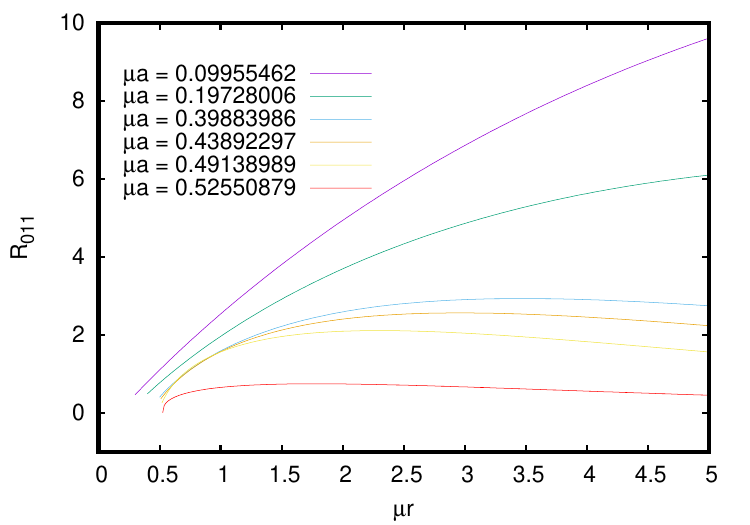}
    \includegraphics[width=0.5\textwidth]{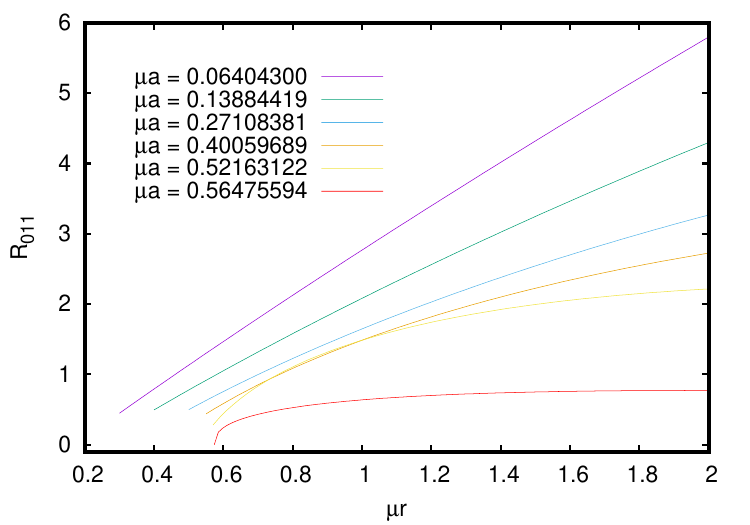}
    \caption{Behavior of the radial functions $R_{nlm}$ that appears in the Fig.~\ref{fig:RadialNewFunction}, but depicted close to the event horizon.}\label{fig:NHRadialNewFunction}
\end{figure}

Figure~\ref{fig:NHRadialNewFunction} shows the radial functions $R_{nlm}$ that are represented in Fig.~\ref{fig:RadialNewFunction} but focusing on their behavior near the event horizon. We see how the value of the radial function on the horizon, $R(r_H)$, changes depending on the location of $r_H$, unlike what was presented in Sec.~\ref{Teukolsky}, where all cloud configurations for the Kerr-Newman scenario have a fixed value $R(r_H) = 1$ (see Figure~\ref{fig:RadialKN3}). For the Kerr scenario see Fig.~3 of Ref.\cite{Garcia2019}.

\subsection{Configurations for $m = l \geq 3$}
\label{subsec:mlarge}

The (uncharged and charged) scalar cloud configurations around extremal Kerr and KN BH's
presented in Sections \ref{subsec:Kerr} and \ref{subsec:KerrNewman}, respectively, correspond to values of $m = l = 1, 2$ only. This is because numerically we observe that for $m = l \geq 3$ the values of the exponent $\alpha_+$ appearing in the radial function (\ref{ansatzExt}) become negative in both scenarios, which immediately implies that the field $\Psi$ becomes divergent at the event horizon including their radial derivatives.

Something to note is that $\alpha_+ \equiv 0$ when
\begin{equation}
  \label{KonstantsKerr}
K_{lm} =  2m^2 -2\mu^2M^2\;,
\end{equation}
for the extremal Kerr scenario [cf. Eq.(\ref{alphaKerr2})], and
\begin{eqnarray}
\nonumber K_{lm} &=& \frac{2m^2a^2 + 2qmaQM}{M^2 + a^2} + \frac{\left(2amM + qQ^3\right)^2}{\left(M^2 + a^2\right)^2}\\
&-& \mu^2\left(M^2 + a^2\right)\;,\label{KonstantsKN}
\end{eqnarray}
for a KNBH [cf. Eq.~(\ref{alphaKN2})]\footnote{When $Q = 0$ in (\ref{KonstantsKN}) the constants $K_{lm}$ reduce to those
of Eq.~(\ref{KonstantsKerr}), where $a = M$ for an extremal Kerr BH.}.
Focusing on the case of an extremal Kerr BH (\ref{metricKerrExt}) and based on the study presented in \cite{Garcia2020}
we see that if we take the separation constants as in Eq.(\ref{KonstantsKerr}), one obtains radial solutions and its
derivatives that are well behaved on the horizon but that diverge on the axis of symmetry.
Something completely analogous happens when one studies the Kerr-Newman scenario and chooses the separation constants that appear in (\ref{KonstantsKN}). It is worth mentioning that this choice of the separation constants arises precisely
when {\it superregularity} conditions are imposed on the radial functions in extremal scenarios \cite{Garcia2020},
but then the separation constants (which incidentally do not depend explicitly on the number $l$) given
by (\ref{KonstantsKerr}) and (\ref{KonstantsKN}) do not coincide in general with the actual separation constants
(\ref{KlmKN}) which makes the spheroidal harmonics to be well behaved in the axis of symmetry\footnote{See \cite{Garcia2020} for more details.}. Thus, we must disregard this possibility and look for
$\alpha_+ >0$.

To understand what happens when $m \geq 3$, we analyze first the following term associated with the Kerr scenario
\begin{equation}
  \label{Theta1}
\Theta \equiv 1 + 4\left[K_{lm} + 2\left(\mu^2M^2 - m^2\right)\right]\;,
\end{equation}
which appears as subradical in Eq.(\ref{alphaKerr2}). Using Eq.~(\ref{KlmKN}) in the previous equation we find
\begin{equation}
  \label{Theta2}
\Theta = 4\left(l + \frac{1}{2}\right)^2 + 4\mu^2M^2 - 7m^2 + 4\lambda\;, 
\end{equation}
where
\begin{equation}\label{lambdaKerr}
\lambda \equiv \sum_{k=1}^{\infty}c_{k}(\mu^2M^2 - m^2/4)^k\;.
\end{equation}
Table~\ref{tab:CloudsKm3} shows some values of certain quantities associated with the parameter $\Theta$ for the
optimal values of $\mu M$ that we found for the extremal Kerr BH leading to asymptotically vanishing clouds.
The table includes the values $m=l=3$ alluded above leading to bad behaved clouds at the horizon ($\alpha_+ <0$),
and also some values $m<l$ leading to well behaved clouds everywhere in the DOC.

\begin{table}[h!]
\centering
\begin{tabular}{|c|c|c|c|c|c|}
\hline
 $m$ & $l$ & $\mu M$ & $\sqrt{\Theta}$ & $\lambda$ & $\alpha_+$\\
\hline \hline
\multirow{3}{0.2cm}{1} & 1 & 0.52550879 & 1.76792337 & 0.00522877 & 0.38396168\\ \cline{2-6}
& 2 & 0.50806701 & 4.36422594 & 0.00348492 & 1.68211297\\ \cline{2-6}
& 3 & 0.50420923 & 6.55932909 & 0.00197260 & 2.77966454\\ \hline
\multirow{3}{0.2cm}{2} & 2 & 1.14093944 & 1.54230077 & 0.04293009 & 0.27115039\\ \cline{2-6}
& 3 & 1.04424934 & 5.04800576 & 0.03013387 & 2.02400474 \\ \cline{2-6}
& 4 & 1.02432966 & 7.56811183 & 0.01982790 & 3.28405591\\ \hline
\multirow{3}{0.2cm}{3} & 3 & 1.87316979 & 0.76699072 & 0.13830364 & -0.11650463\\ \cline{2-6}
& 4 & 1.61385150 & 5.36694663 & 0.09651237 & 2.18347331\\ \hline
\end{tabular}
\caption{Quantities $\mu M$, $\Theta$ and $\lambda$ that contributes to the parameter $\alpha_+$
  given in Eq.(\ref{alphaKerr2}). These values correspond to $n = 0$. Clouds are not regular when
$\alpha_+<0$, namely, when $m=l=3$ as one notices in the second-last row.}
\label{tab:CloudsKm3}
\end{table}

In \cite{Hod2012} Hod finds that the existence of \textit{bound states} implies that\footnote{See Eq.~(18) in \cite{Hod2012}.}
\begin{equation}\label{bandKerr}
\frac{m}{2} < \mu M < \frac{m}{\sqrt{2}}\;.
\end{equation}
Using the inequalities (\ref{bandKerr}) in $\Theta$ given by Eq.(\ref{Theta2}) it is possible
to observe the corresponding inequalities:
\begin{equation}\label{ThetaMm}
\Theta_{\rm min} < \Theta < \Theta_{\rm max}\;,
\end{equation}
where\footnote{For $m = l$ we have that $\Theta_{\rm min} = 2m(2 - m) + 1$.}
\begin{equation}\label{Thmin}
\Theta_{\rm min} = 4\left(l + \frac{1}{2}\right)^2 - 6m^2 \;,
\end{equation}
and
\begin{equation}\label{Thmax}
\Theta_{\rm max} = 4\left(l + \frac{1}{2}\right)^2 - 5m^2 + \sum_{k=1}^{\infty}c_{k}\left(\frac{m^2}{4}\right)^k\;.
\end{equation}
Now, in order for $\alpha_+>0$, we require $\Theta >1.$ If $\Theta$ saturates the lower
  bound of (\ref{ThetaMm}) and in addition $\Theta_{\rm min} \geq 1$, then these conditions are sufficient for $\alpha_+>0$.
  From Eq.(\ref{Thmin}) we find that $1 \leq \Theta_{\rm min}$ if
\begin{equation}
|m| \leq \sqrt{\frac{2l(l+1)}{3}}\;. 
\end{equation}
For example, taking $l = 1$, $|m|< \sqrt{4/3} \approx 1.15$; for $l = 2$, $|m| \leq 2$; for $l = 3$, $|m|\leq\sqrt{8} \approx 2.83$; and for $l = 4$, $|m|\leq\sqrt{40/3} \approx 3.65$, and so on\footnote{For $m = l$, $\Theta_{\rm min} \geq 1$ implies $2m(2 - m) \geq 0$ and then $0 \leq m \leq 2$.}. In the case $l=3$, and given that $m$ is an integer, the condition
$|m|\leq\sqrt{8}$ can be only satisfied if for instance, $|m|= 2$, but not for $|m|=3$. This trend is confirmed by
the numerical analysis as shown in Table~\ref{tab:CloudsKm3}. From the table we appreciate that for
$l=3=m$, $\sqrt{\Theta}< 1$, and thus $\alpha_+ <0$, which leads to singular scalar clouds at the horizon.
This problem persists for $l \geq 4 $ which forces $|m|< l$ when $l\geq 3$. Notwithstanding, when $|m|< l$, the numerical
evidence shows that $\alpha_+ > 1$, which leads not only to a field $\Psi$ that vanishes at the horizon, but its
radial derivative also vanishes there (cf. Eq.(\ref{Rpextremal})).

We thus conjecture that if $m = l \geq 3$ the scalar cloud configurations are not regular at the horizon.
On the other hand, when $|m| < l$ the scalar
clouds around an extremal Kerr BH are regular everywhere and the field and its radial derivative $R'$ vanish
at the horizon $r=M$ since in this situation $\alpha_+ > 1$. Even $R''$ vanish at the horizon when $\alpha_+ > 2$.
A similar behavior occurs in extremal KNBH
(see Table~\ref{tab:CloudsKNm3}) and we extend our conjecture to that scenario as well.
\begin{table}[h!]
\centering
\begin{tabular}{|c|c|c|c|c|}
\hline
 $l$ & $m$ & $\mu a$ & $\mu M$ & $\alpha_+$\\
\hline \hline
3 & 2 & 1.08761968 & 1.09220720 & 2.02617256\\ \hline
3 & 3 & 1.90941874 & 1.91203554 & -0.11159253\\ \hline
4 & 3 & 1.65721537 & 1.66022974 & 2.18544718\\ \hline
\end{tabular}
\caption{Behavior of the exponent $\alpha_+$ given in Eq.(\ref{alphaKN2}) for some values of the parameters
  ($n = 0$, $\mu Q = 0.1$ and $q/\mu = 1$) leading to charged clouds around a KNBH. Only the clouds with
  $\alpha_+>0$ are regular in the DOC, including the horizon.}
\label{tab:CloudsKNm3}
\end{table}

\begin{figure}[h!]
   \includegraphics[width=0.5\textwidth]{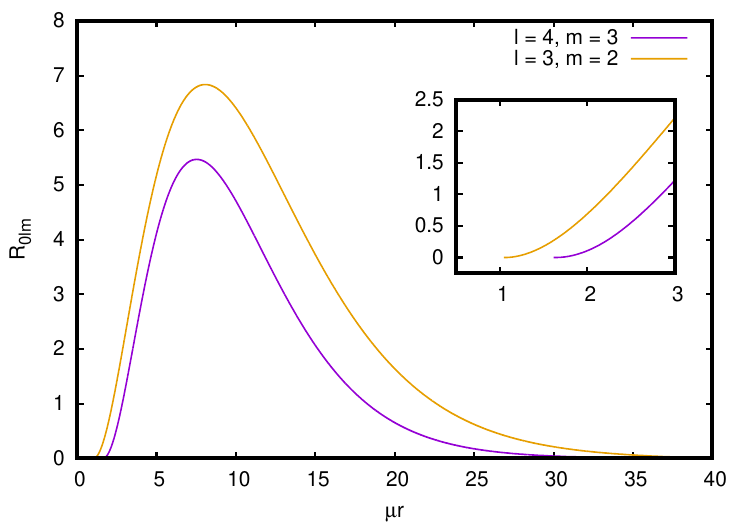}
    \includegraphics[width=0.5\textwidth]{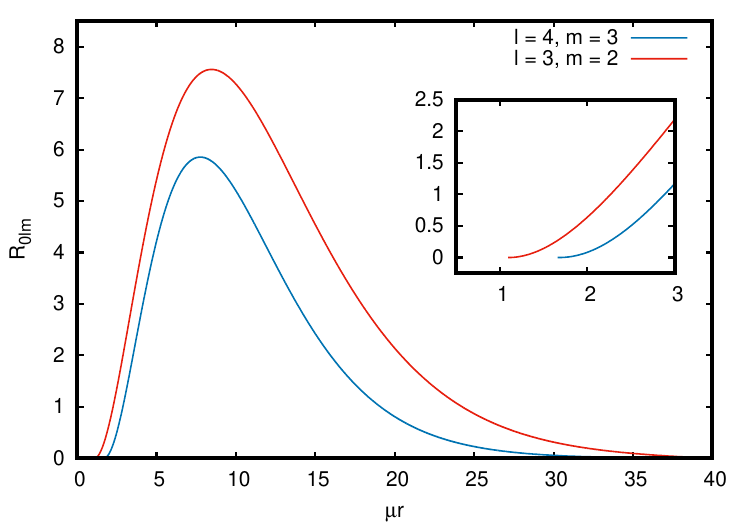}
\caption{Radial part $R_{nlm}$ of the scalar field $\Psi$ associated with integer numbers $n = 0$ and $l \neq m$, considering $l = \{4, 3\}$ and $m = \{3, 2\}$, respectively, around an extremal Kerr BH (top panel) and an extremal Kerr-Newman BH (bottom panel) with $\mu Q = 0.1$.}\label{fig:Radiallneqm}
\end{figure}

Figure~\ref{fig:Radiallneqm} depicts some solutions for the radial function $R(r)$ with  $m <l$,
around extremal Kerr BH's (top panel) and extremal KNBH (bottom panel). These radial functions
have $\alpha_+>0$, and thus, provide clouds that are regular in the DOC. The plots
  confirm that $R(M)=0= R'(M)$.

In this regard, it is important to emphasize that our parametrization (\ref{ansatzExt}) allows us to find numerically
nontrivial regular clouds even in cases where $R(M)=0= R'(M)= R''(M)$ as otherwise treating directly the
differential equation for $R(r)$ with those conditions at the horizon would not necessarily lead to the
actual non-trivial solution but rather to the trivial one $R(r)\equiv 0$.

\section{Conclusion}
\label{sec:conslusion}

We analyzed the existence of scalar clouds, electrically charged and non-charged, in the backgrounds of Kerr and Kerr-Newman
BH's (extremal and subextremal), respectively.  The electrically charged cloud solutions presented in this work
are consistent with those reported in Ref.~\cite{Benone2014}. Some of the results presented in \cite{Garcia2019}
about the existence of clouds around a Kerr BH's were extended here as to include the electric charge in both the clouds and
the BH by establishing an integral method that allows us to understand and justify in a simple and heuristic way the existence of
such non-trivial cloud solutions in this type of scenario. Since a similar method is precisely used to prove no-hair theorems
in static spacetimes, we illustrate that when rotating BH's are present there exist obstructions to extend those theorems
due to the presence of non-positive definite terms associated with the rotation which does {\it not} allow us to conclude
that only trivial fields are possible around rotating BH's.


Moreover, a new technique based on factorizing the radial function in terms of a regular part and a potentially irregular one
at the horizon of the form $(r-M)^\alpha$, allows us to analyze clouds around {\it exact} extremal Kerr and KN BH's. In this way we can analyze the exponent $\alpha$ and when $\alpha>0$ the cloud solutions turn to be well behaved in the DOC (including the horizon), where the scalar field associated with the clouds vanishes. Remarkably even if the radial derivatives are unbounded at the horizon when $0<\alpha <1$, the cloud solutions are still regular since the covariant (coordinate invariant) kinetic term that contains radial derivatives turns to be bounded due to the presence of an (inverse) metric component that vanishes at the horizon. This treatment closes a gap left in a previous analysis by us  \cite{Garcia2020} where boundedness of radial derivatives at the
horizon (superregularity conditions) were imposed leading to some inconsistencies. Furthermore, we checked that
the cloud solutions without the superregularity restriction are in excellent agreement with the exact solutions
found in the past by Hod \cite{Hod2012,Hod2015}. Finally, the factorization technique made possible to find cloud solutions
when $m < l$, even for large $l$ or when $m= l$ for $l=1,2$ while providing us with some insight about the impossibility to
find regular solutions when $l = m \geq 3$ which are associated with $\alpha <0$. It is possible that a
similar factorization technique might be implemented when dealing with hairy-rotating black holes (at the full non-linear level)
like those analyzed recently in \cite{Garcia2023}, which may help us to deal with the extremal scenario in quasi-isotropic
coordinates (as opposed to the Boyer-Lindquist coordinates used in this work)
which are inherently singular at the horizon located at $r_{\rm qiso}=0$ \cite{Garcia2023,Garcia2023b}.


\section*{Acknowledgments}
This work was supported partially by DGAPA--UNAM grant IN105223 and CONACYT (FORDECYT-PRONACES) grant 140630. We are indebted to P. Grandcl\'ement and E. Gourgoulhon for their helpful comments and discussions.    


\appendix
\section{Data Kerr-Newman-Bosonic field}
\label{sec:appendixKN}
The following Tables~\ref{tab:spectrum1KN}-\ref{tab:spectrum3KN} show the eigenvalues $\mu a$ found for different values of $\mu r_{H}$ that leads to cloud solutions associated with a massive and charged scalar field around Kerr-Newman black holes
with electric charge $\mu Q = 0.1$. The numerical data corresponds to the {\it fundamental mode} with $n = 0$ (nodeless).
\begin{table*}[h]
\begin{center}
\begin{tabular}{|c|c|c|c|c|}
\hline
$\boldsymbol{\mu r_H}$ & $\boldsymbol{\mu a}$ & $\boldsymbol{\omega/\mu}$ & $\boldsymbol{\Omega_H/\mu}$ & $\boldsymbol{\mu M}$\\
\hline 
0.12 &    0.0024057827214  &   0.9999996481207  &   0.1670011217984  &  0.1016907824604  \\ \hline         
0.13 &    0.0039153033832  &   0.9999984500873  &   0.2314648000216  &  0.1035204984637  \\ \hline        
0.14 &    0.0056316381056  &   0.9999960654455  &   0.2868642910689  &   0.1058275548134 \\ \hline    
0.15 &    0.0075569016729  &   0.9999909079654  &   0.3350120098438  &   0.1085236892096 \\ \hline        
0.16 &    0.0096935363204  &   0.9999833297078  &   0.3772689989226  &   0.1115436395199 \\ \hline    
0.18 &    0.0146121358341  &   0.9999577267443  &   0.4480392883781  &   0.1183708736489 \\ \hline    
0.20 &    0.0204132614973  &   0.9999148929166  &   0.5050699554542  &   0.1260417531124 \\ \hline   
0.22 &    0.0271286580352  &   0.9998513094893  &   0.5521140887208  &   0.1343999183790 \\ \hline    
0.24 &    0.0347969078252  &   0.9997633604195  &   0.5916752221561  &   0.1433558849879 \\ \hline    
0.26 &    0.0434646737536  &   0.9996473679561  &   0.6254883866728  &   0.1528638035856 \\ \hline    
0.28 &    0.0531883296210  &   0.9994993316724  &   0.6547948474511  &   0.1629089257283 \\ \hline    
0.30 &    0.0640361280605  &   0.9993145886544  &   0.6805069316617  &   0.1735010428283 \\ \hline    
0.32 &    0.0760911342624  &   0.9990874703943  &   0.7033111398859  &   0.1846716573645 \\ \hline    
0.34 &    0.0894552748854  &   0.9988109332642  &   0.7237350261189  &   0.1964738914776 \\ \hline    
0.36 &    0.1042551145940  &   0.9984764316070  &   0.7421923620962  &   0.2089849012764 \\ \hline    
0.38 &    0.1206497816675  &   0.9980712433904  &   0.7590119339466  &   0.2223110129163 \\ \hline    
0.40 &    0.1388439245643  &   0.9975807867885  &   0.7744631630351  &   0.2365970442355 \\ \hline    
0.42 &    0.1591062367041  &   0.9969830777394  &   0.7887683055310  &   0.2520414220930 \\ \hline    
0.44 &    0.1818009833365  &   0.9962470407675  &   0.8021164876315  &   0.2689222699342 \\ \hline    
0.45 &    0.1942129892664  &   0.9958134150449  &   0.8084841073244  &   0.2780207613331 \\ \hline    
0.46 &    0.2074420085504  &   0.9953258919251  &   0.8146731607918  &   0.2876436814255 \\ \hline    
0.48 &    0.2367940433120  &   0.9941447016175  &   0.8265887193270  &   0.3088243947375 \\ \hline    
0.50 &    0.2710838133913  &   0.9925727318373  &   0.8380067446339  &   0.3334864338828 \\ \hline    
0.52 &    0.3125126793142  &   0.9903508799232  &   0.8490711695634  &   0.3635232449347 \\ \hline    
0.54 &    0.3658458319769  &   0.9868435054170  &   0.8599170356659  &   0.4031881229396 \\ \hline    
0.55 &    0.4005968950143  &   0.9840576024849  &   0.8652614282156  &   0.4299798839046 \\ \hline      
0.56 &    0.4464066171732  &   0.9795814346444  &   0.8703938593518  &   0.4668561320143 \\ \hline      
0.57 &    0.5216314025034  &   0.9692329338467  &   0.8737554381729  &   0.5324555439278 \\ \hline    
0.571 &   0.5331035527382  &   0.9671656526322  &   0.8735959705977  &   0.5431176864642 \\ \hline     
0.572 &  0.5456459308501   &   0.9646870300545  &   0.8731543592753  &   0.5549943023193 \\ \hline    
0.573 &   0.5589214827882  &   0.9617607440179  &   0.8723304138988  &   0.5678204397227 \\ \hline  
\end{tabular}
\caption{Eigenvalues associated with charged scalar clouds $(q/\mu = 1)$ around a Kerr-Newman black hole $(\mu Q = 0.1)$ with parameters $l = m = 1$, which are indicated by the dotted line labeled $m=1$ in Figure~\ref{fig:RadialKN1}.}
\label{tab:spectrum1KN}
\end{center}
\end{table*}

\begin{table*}[h]
\begin{center}
\begin{tabular}{|c|c|c|c|c|}
\hline
$\boldsymbol{\mu r_H}$ & $\boldsymbol{\mu a}$ & $\boldsymbol{\omega/\mu}$ & $\boldsymbol{\Omega_H/\mu}$ & $\boldsymbol{\mu M}$\\
\hline 
0.11  &   0.0005501512807  &   0.9999999913311  &   0.0454659107242  &   0.1004559212110 \\ \hline        
0.12  &   0.0012007197534  &   0.9999998457823  &   0.0833749687127  &   0.1016726738663 \\ \hline        
0.13  &   0.0019518996880  &   0.9999993766910  &   0.1154709913424  &   0.1034761919707 \\ \hline  
0.14  &   0.0028039130146  &   0.9999981691570  &   0.1429994267048  &   0.1057423640292 \\ \hline    
0.15  &   0.0037570136356  &   0.9999960966621  &   0.1668736972497  &   0.1083803838381 \\ \hline         
0.20  &   0.0100497880790  &   0.9999645374804  &   0.2506119179064  &   0.1252524956010 \\ \hline    
0.25  &   0.0189254314487  &   0.9998837214336  &   0.3010814800992  &   0.1457163439110 \\ \hline    
0.30  &   0.0304518478210  &   0.9997402387422  &   0.3349031895315  &   0.1682121917261 \\ \hline    
0.35  &   0.0447203722441  &   0.9995247610664  &   0.3592000318380  &   0.1921427309909 \\ \hline    
0.40  &   0.0618494070021  &   0.9992276217155  &   0.3775325816450  &   0.2172816864331 \\ \hline    
0.45  &   0.0819895860755  &   0.9988379819796  &   0.3918778692444  &   0.2435803246942 \\ \hline    
0.50  &   0.1053309163894  &   0.9983424813634  &   0.4034205058387  &   0.2710946019474 \\ \hline    
0.53  &   0.1209694101014  &   0.9979874437927  &   0.4093254061340  &   0.2882392435663 \\ \hline    
0.56  &   0.1379073613568  &   0.9975836649289  &   0.4146112922233  &   0.3059093217110 \\ \hline    
0.59  &   0.1562164495946  &   0.9971257190680  &   0.4193689895868  &   0.3241555755287 \\ \hline    
0.62  &   0.1759797916451  &   0.9966071031775  &   0.4236710960409  &   0.3430394250544 \\ \hline    
0.65  &   0.1972942730077  &   0.9960199728535  &   0.4275760093436  &   0.3626346385859 \\ \hline     
0.68  &   0.2202735266691  &   0.9953547106563  &   0.4311307891055  &   0.3830297254053 \\ \hline    
0.71  &   0.2450518675579  &   0.9945995206569  &   0.4343732802970  &   0.4043312801363 \\ \hline    
0.74  &   0.2717894364571  &   0.9937396366685  &   0.4373335094200  &   0.4266685795741 \\ \hline    
0.77  &   0.3006791637884  &   0.9927563671831  &   0.4400346280063  &   0.4501999737249 \\ \hline    
0.80  &   0.3319562973226  &   0.9916256588933  &   0.4424933579909  &   0.4751218645825 \\ \hline    
0.83  &   0.3659117265932  &   0.9903159676282  &   0.4447199257343  &   0.5016815612400 \\ \hline    
0.86  &   0.4029110404802  &   0.9887850014551  &   0.4467173467138  &   0.5301961084539 \\ \hline    
0.89  &   0.4434224612577  &   0.9869745617414  &   0.4484797525389  &   0.5610806062628 \\ \hline    
0.92  &   0.4880588597134  &   0.9848020384726  &   0.4499891268523  &   0.5948920926873 \\ \hline    
0.95  &   0.5376425365021  &   0.9821457307799  &   0.4512091404838  &   0.6323997352928 \\ \hline    
0.98  &   0.5933068329668  &   0.9788181524010  &   0.4520732679808  &   0.6747005092067 \\ \hline    
1.01  &   0.6566536849725  &   0.9745146810617  &   0.4524608087165  &   0.7234129019742 \\ \hline    
1.04  &   0.7299691681541  &   0.9687097809656  &   0.4521458785027  &   0.7809879742575 \\ \hline    
1.07  &   0.8163174734726  &   0.9604481365331  &   0.4506868510505  &   0.8510627184563 \\ \hline    
1.10  &   0.9182234019219  &   0.9480368057917  &   0.4472300908723  &   0.9377882799259 \\ \hline    
1.13  &   1.0301793052790  &   0.9295128956888  &   0.4405922448678  &   1.0390130093032 \\ \hline    
1.16  &   1.1292776607708  &   0.9060186509682  &   0.4308792528429  &   1.1339948427224 \\ \hline    
1.17  &   1.1554073083968   &   0.8979049159343  &   0.4273167707862  &   1.1597718155114 \\ \hline     
1.171 &    1.1577936094314  &   0.8970988529260  &   0.4269580202881  &   1.1621379342614 \\ \hline   
1.172 &    1.1601361476599  &   0.8962943195009  &   0.4265990801174  &   1.1644624066156 \\ \hline   
1.173 &    1.1624338345399  &   0.8954914592236  &   0.4262400096121  &   1.1667439981599 \\ \hline   
1.174 &    1.1646852618309  &   0.8946904231618  &   0.4258808660970  &   1.1689811580605 \\ \hline   
1.175 &    1.1668885219775  &   0.8938913763719  &   0.4255217049233  &   1.1711718394565 \\ \hline   
1.176 &    1.1690408371188  &   0.8930945110393  &   0.4251625794030  &   1.1733131287633 \\ \hline   
1.177 &    1.1711375085258  &   0.8923000828668  &   0.4248035401426  &   1.1754001970586 \\ \hline   
1.178 &    1.1731655545607  &   0.8915086268446  &   0.4244446282652  &   1.1774199568793 \\ \hline    
\end{tabular}
\caption{Similar to Table~\ref{tab:spectrum1KN} but for the mode $l = m = 2$ with $n = 0$. These eigenvalues are associated with the dashed line labeled $m=2$ in Figure~\ref{fig:RadialKN1}.}
\label{tab:spectrum2KN}
\end{center}
\end{table*}

\begin{table*}[h]
\begin{center}
\begin{tabular}{|c|c|c|c|c|}
\hline
$\boldsymbol{\mu r_H}$ & $\boldsymbol{\mu a}$ & $\boldsymbol{\omega/\mu}$ & $\boldsymbol{\Omega_H/\mu}$ & $\boldsymbol{\mu M}$\\
\hline 
0.11  &  0.0003667114669  &  0.9999999936894  &   0.0303063959882  &   0.1004551567150 \\ \hline         
0.12  &  0.0008002130296  &  0.9999999130702  &   0.0555678782846  &   0.1016693347537 \\ \hline        
0.13  &  0.0013005618428  &  0.9999996490018  &   0.0769486205492  &   0.1034680440811 \\ \hline         
0.14  &  0.0018678229089  &  0.9999989779648  &   0.0952801275132  &   0.1057267455800 \\ \hline        
0.15  &  0.0025020704490  &  0.9999978224055  &   0.1111721987292  &   0.1083542011884 \\ \hline        
0.20  &  0.0066812847781  &  0.9999803910845  &   0.1668459212137  &   0.1251115989157 \\ \hline   
0.25  &  0.0125511715278  &  0.9999358844349  &   0.2003138512792  &   0.1453150638134 \\ \hline   
0.30  &  0.0201308031580  &  0.9998579754837  &   0.2226729457434  &   0.1673420820596 \\ \hline   
0.35  &  0.0294451011920  &  0.9997429587765  &   0.2386788837717  &   0.1905243056917 \\ \hline   
0.40  &  0.0405252189993  &  0.9995877540750  &   0.2507092553147  &   0.2145528667186 \\ \hline   
0.45  &  0.0534090512068  &  0.9993894736426  &   0.2600847043470  &   0.2392805852786 \\ \hline   
0.50  &  0.0681418718515  &  0.9991450820340  &   0.2675973328890  &   0.2646433146994 \\ \hline    
0.55  &  0.0847771353082  &  0.9988512383161  &   0.2737508864655  &   0.2906246933373 \\ \hline    
0.60  &  0.1033774554613  &  0.9985040423163  &   0.2788807911368  &   0.3172390819147 \\ \hline    
0.65  &  0.1240158270189  &  0.9980989210825  &   0.2832188000384  &   0.3445230195009 \\ \hline    
0.70  &  0.1467771341929  &  0.9976304566891  &   0.2869299021704  &   0.3725310908013 \\ \hline    
0.75  &  0.1717600232967  &  0.9970921767998  &   0.2901344366039  &   0.4013343370686 \\ \hline    
0.80  &  0.1990792424803  &  0.9964763056687  &   0.2929218796355  &   0.4310203404916 \\ \hline    
0.85  &  0.2288685825797  &  0.9957734399481  &   0.2953597176268  &   0.4616946047600 \\ \hline    
0.90  &  0.2612846044529  &  0.9949721236604  &   0.2974992999951  &   0.4934831358467 \\ \hline    
0.95  &  0.2965114078248  &  0.9940582809832  &   0.2993797608315  &   0.5265363236685 \\ \hline    
1.00  &  0.3347667970329  &  0.9930144454746  &   0.3010306507165  &   0.5610344041978 \\ \hline    
1.05  &  0.3763103452734  &  0.9918186941452  &   0.3024736589061  &   0.5971949885522 \\ \hline    
1.10  &  0.4214540724942  &  0.9904431444100  &   0.3037236410283  &   0.6352834251008 \\ \hline    
1.15  &  0.4705767659242  &  0.9888517901817  &   0.3047890502212  &   0.6756271707076 \\ \hline    
1.20  &  0.5241434279212  &  0.9869973133089  &   0.3056717668726  &   0.7186359720970 \\ \hline    
1.25  &  0.5827319765302  &  0.9848162647386  &   0.3063662051602  &   0.7648306225883 \\ \hline    
1.30  &  0.6470701365174  &  0.9822215790490  &   0.3068574048848  &   0.8148845236817 \\ \hline    
1.35  &  0.7180861622466  &  0.9790906071036  &   0.3071175319952  &   0.8696843468185 \\ \hline    
1.40  &  0.7969763040270  &  0.9752454418509  &   0.3070996992667  &   0.9304182961359 \\ \hline    
1.45  &  0.8852846376814  &  0.9704198870107  &   0.3067271070690  &   0.9986996171429 \\ \hline    
1.50  &  0.9849584254600  &  0.9642041300870  &   0.3058741164314  &   1.0767143666282 \\ \hline    
1.55  &  1.0982215461176  &  0.9559590030595  &   0.3043353150022  &   1.1672872788248 \\ \hline    
1.60  &  1.2267292840044  &  0.9447270945530  &   0.3017884637265  &   1.2733952300731 \\ \hline    
1.65  &  1.3687597323441  &  0.9293464357228  &   0.2978152258224  &   1.3957585469354 \\ \hline    
1.70  &  1.5144783472102  &  0.9092897655990  &   0.2921647692556  &   1.5275425482848 \\ \hline    
1.75  &  1.6474738004273  &  0.8858774119710  &   0.2851943805618  &   1.6533342637413 \\ \hline    
1.80  &  1.7569780179411  &  0.8615391793175  &   0.2776965167271  &   1.7602699320911 \\ \hline    
1.85  &  1.8416061126351  &  0.8379523602049  &   0.2702674756578  &   1.8443278578635 \\ \hline    
1.86  &  1.8554705618497  &  0.8333964779116  &   0.2688164052998  &   1.8581642488953  \\ \hline   
1.861  &  1.8567869929913  &  0.8329443673848  &    0.2686720665517  &   1.8594784893448 \\ \hline   
1.862  & 1.8580878534479   &  0.8324929349477  &    0.2685278700756  &   1.8607772478868 \\ \hline  
1.863  &  1.8593800880717  &  0.8320420880535  &    0.2683838187323  &   1.8620674481796 \\ \hline  
1.864  &  1.8606487211222  &  0.8315920316224  &    0.2682399090332  &   1.8633341371817 \\ \hline    
1.865  &  1.8618953389739  &   0.8311427465329   &  0.2680961426494  &   1.8645788882822 \\ \hline    
1.866  &  1.8631094879049  &   0.8306943781580   &  0.2679525183212  &   1.8657912550700 \\ \hline    
\end{tabular}
\caption{Similar to Table~\ref{tab:spectrum1KN} but for the mode $l = m = 3$ with $n = 0$. These eigenvalues are associated with the dotted line labeled $m=3$ in Figure~\ref{fig:RadialKN1}.}
\label{tab:spectrum3KN}
\end{center}
\end{table*}



\begin{thebibliography}{}

\bibitem{LIGOVirgo}
LIGO Scientific Collaboration \& Virgo Collaboration, \Journal{\PRL}{116}{061102}{2019}.

\bibitem{EHTM87} The Event Horizon Telescope Collaboration et. al., \Journal{\PRD}{875}{L1-L5}{2019}.

\bibitem{EHTSagA}
The Event Horizon Telescope Collaboration et. al., \Journal{\PRD}{930}{L12}{2022}.

\bibitem{Carter1968}
B. Carter, \Journal{\PR}{174}{1559}{1968}.

\bibitem{Ruffini1971}
R. Ruffini and J. A. Wheeler, Phys. Today {\bf 24}(1), 30 (1971).


\bibitem{Uniqueness}
W. Israel, \Journal{\PR}{164}{1776}{1967}; \Journal{\CMP}{8}{245}{1971}; B.
Carter, \Journal{\PRL}{26}{331}{1971}; R. Wald,
\Journal{\ibid}{26}{1653}{1971}; D. C. Robinson,
\Journal{\ibid}{34}{905}{1977}; P. Q. Mazur, \Journal{\JPA}{15}{3173}{1982};
\Journal{\PLA}{100}{341}{1984}.

\bibitem{HeuslerBook}
M. Heusler, {\it Black Hole Uniqueness Theorems}, Cambridge University Press, 1996.

\bibitem{Bekenstein1972}
J. D. Bekenstein, \Journal{\PRL}{28}{452}{1972}; {\it idem}
\Journal{\PRD}{5}{1239}{1972}; {\it idem} \Journal{\PRD}{5}{2403}{1972}.

\bibitem{Bekenstein1995}
J. D. Bekenstein, \Journal{\PRD}{51}{R6608}{1995}.

\bibitem{Sudarsky1995}
D. Sudarsky, \Journal{\CQG}{12}{579}{1995}.

\bibitem{Pena1997} 
I. Pe\~na, and D. Sudarsky, \Journal{\CQG}{14}{3131}{1997}.

\bibitem{Sudarsky1998}
D. Sudarsky, and T. Zannias, \Journal{\PRD}{58}{087502}{1998}.

\bibitem{Hod2012}
S. Hod, \Journal{\PRD}{86}{104026}{2012}; \Journal{\PRD}{86}{129902(E)}{2012}.

\bibitem{Hod2013}
S. Hod, \Journal{\EPJC}{73}{2378}{2013}.

\bibitem{Hod2014}
S. Hod, \Journal{\PRD}{90}{024051}{2014}

\bibitem{Hod2015}
S. Hod, \Journal{\PLB}{751}{177-183}{2015}.

\bibitem{Herdeiro2014}
C. Herdeiro, and E. Radu, \Journal{\PRL}{112}{221101}{2014}.

\bibitem{Herdeiro2015}
C. Herdeiro, and E. Radu,\Journal{\CQG}{32}{144001}{2015}.

\bibitem{Benone2014}
C. Benone, L. C. B. Crispino, C. Herdeiro, and E. Radu, \Journal{\PRD}{90}{104024}{2014}.

\bibitem{Grandclement2022}
P. Grandcl\'ement, and  Jordan Nicoules, \Journal{\PRD}{105}{104011}{2022}.
 
\bibitem{Garcia2023}
G. Garc\'ia, E. Gourgoulhon, P. Grandcl\'ement and M. Salgado, \Journal{\PRD}{107}{084047}{2023}.

\bibitem{Delgado2016}
J. F. M. Delgado, C. A. R. Herdeiro, E. Radu, and H. R\'unarsson \Journal{\PLB}{761}{234}{2016}.

\bibitem{Garcia2019}
G. Garc\'ia and M. Salgado, \Journal{\PRD}{99}{044036}{2019}.

\bibitem{Garcia2021}
G. Garc\'ia and M. Salgado, \Journal{\PRD}{104}{064054}{2021}.

\bibitem{Garcia2020}
G. Garc\'ia and M. Salgado, \Journal{\PRD}{101}{044040}{2020}.

\bibitem{Hawking1976}
S. W. Hawking, \Journal{\PRD}{191}{044040}{1976}.

\bibitem{Heusler1996}
M. Heuler, \Journal{\HPA}{69}{501}{1996}
  
\bibitem{Ayon2002}
E. Ay\'on-Beato, \Journal{\CQG}{19}{5465}{2002}

\bibitem{Abramowitz}
M. Abramowitz and I.A. Stegun, {\it Handbook of Mathematical Functions with Formulas, Graps, and Mathematical Tables}, Dover, 1964.

\bibitem{Teukolsky1972}
S. A. Teukolsky, \Journal{\PRL}{29}{1114}{1972}.

\bibitem{Hong2020}
J.P Hong, M. Suzuki, and M. Yamada, \Journal{\PLB}{803}{135324}{2020}.

\bibitem{Herdeiro2020}
C.A.R. Herdeiro, and E. Radu, \Journal{\EPJC}{80}{390}{2020}.
    
\bibitem{Brihaye2021}
Y. Brihaye and B. Hartmann, \Journal{\CQG}{38}{06LT01}{2021}.

\bibitem{DegolladoHerdeiro2013}
  J. C. Degollado and C. A. R. Herdeiro, \Journal{\GRG}{45}{2483-2492}{2013}.

\bibitem{Garcia2023b}
G. Garc\'ia, E. Gourgoulhon, P. Grandcl\'ement and M. Salgado, in preparation.
  





\end{thebibliography}
\end{document}